\documentclass[a4paper]{article}
\usepackage{amsmath,amssymb,amsfonts,bm,xcolor}

\usepackage{algorithm,algorithmic,url}
\newcommand{\algrule}[1][.2pt]{\par\vskip.5\baselineskip\hrule height #1\par\vskip.5\baselineskip}

\newcommand{\dn}{\textrm{d}}       % Differential for integrals etc.
\newcommand{\dd}{\, \textrm{d}}       % Differential for integrals etc.
\newcommand{\dint}{\int \!}       % Differential for integrals etc.

\DeclareMathOperator*{\argmax}{arg\!\max}

\usepackage{enumitem,hyperref,natbib,multirow,rotating,sectsty,booktabs}           
\allsectionsfont{\sffamily}
\usepackage[margin=1in]{geometry}

\newcommand{\setalglineno}[1]{%
  \setcounter{ALC@line}{\numexpr#1-1}}

\usepackage{setspace}

\title{Bayesian optimisation for fast approximate inference in state-space models with intractable likelihoods}

\author{Johan Dahlin, Mattias Villani and Thomas B.\ Sch\"{o}n%
\thanks{E-mail to corresponding author: \url{liu@johandahlin.com}. JD and MV are with the Department of Computer and Information Science, Link\"{o}ping University. TS is with the Department of Information Technology, Uppsala University, Sweden. This work was mainly carried out while JD worked at the Division of Automatic Control, Department of Electrical Engineering, Link\"{o}ping University.}%
}

\begin{document}
\maketitle

\doublespacing
\begin{abstract}
\noindent We consider the problem of approximate Bayesian parameter inference in non-linear state-space models with intractable likelihoods. Sequential Monte Carlo with approximate Bayesian computations (\textsc{smc}-\textsc{abc}) is one approach to approximate the likelihood in this type of models. However, such approximations can be noisy and computationally costly which hinders efficient implementations using standard methods based on optimisation and Monte Carlo. We propose a computationally efficient novel method based on the combination of Gaussian process optimisation and \textsc{smc}-\textsc{abc} to create a Laplace approximation of the intractable posterior. We exemplify the proposed algorithm for inference in stochastic volatility models with both synthetic and real-world data as well as for estimating the Value-at-Risk for two portfolios using a copula model. We document speed-ups of between one and two orders of magnitude compared to state-of-the-art algorithms for posterior inference. \\ 

\end{abstract}

\newpage
%%%%%%%%%%%%%%%%%%%%%%%%%%%%%%%%%%%%%%%%%%%%%%%%%%%%%%%%%%%%%%%%%%%%%%%%%%%%%%%%%%%%%%%%%%%%%%%%%%%%%%%
%%%%%%%%%%%%%%%%%%%%%%%%%%%%%%%%%%%%%%%%%%%%%%%%%%%%%%%%%%%%%%%%%%%%%%%%%%%%%%%%%%%%%%%%%%%%%%%%%%%%%%%
%%%%%%%%%%%%%%%%%%%%%%%%%%%%%%%%%%%%%%%%%%%%%%%%%%%%%%%%%%%%%%%%%%%%%%%%%%%%%%%%%%%%%%%%%%%%%%%%%%%%%%%
\maketitle

\section{Introduction}
\label{sec:introduction}
% State-space model
Dynamical modelling of time series data is an essential part of many scientific fields including statistics \citep{DurbinKoopman2012}, econometrics \citep{McNeilFreyEmbrechts2010} and engineering \citep{Ljung1999}. A popular dynamical model is the state-space model (\textsc{ssm}) which can be expressed by
\begin{align}
	x_{0} \sim \mu_{\theta}( x_0 ), \quad
	x_{t} | x_{t-1} \sim f_{\theta}( x_{t} | x_{t-1}), \qquad
	y_{t}   | x_t \sim g_{\theta}( y_{t  } | x_t),
	\label{eq:ssmDef}
\end{align}
where $\theta \in \Theta \subseteq \mathbb{R}^p$ denotes unknown static parameters. Here, $x_t \in \mathcal{X} \subseteq \mathbb{R}^{n_x}$ and $y_t \in \mathcal{Y} \subseteq \mathbb{R}^{n_y}$ denotes the latent state and the observations at time $t \in \{0,1,\ldots,T\}$, respectively. The distribution of the initial state, the state dynamics and the observations are modelled by using the known probability densities $\mu$, $f_{\theta}$ and $g_{theta}$, respectively.

% Bayesian parameter inference
In this paper, we are interested in estimating the unknown parameters $\theta$ in \textsc{ssm}s using a Bayesian approach. This amounts to computing the \textit{parameter posterior distribution} given by
\begin{align}
	p( \theta | y_{1:T} )
	=
	\frac{
	p(\theta) p_{\theta}(y_{1:T}) 
	}{
	\displaystyle
	\int_{\Theta} \! p(\theta') p_{\theta'}(y_{1:T}) \dd \theta'
	},
	\label{eq:ParameterPosteriorDef}
\end{align}
where $p(\theta)$ and $p_{\theta}(y_{1:T}) \triangleq p(y_1,y_2,\ldots,y_T|\theta)$ denote the parameter prior distribution and the likelihood, respectively. For an \textsc{ssm}, we cannot compute the posterior in closed-form due to that the likelihood $p_{\theta}(y_{1:T})$ depends on the unknown latent states $x_{0:T}$. Fortunately, it is possible to obtain unbiased estimates of the likelihood via so-called sequential Monte Carlo (\textsc{smc}; \citealp{DoucetJohansen2011}) or particle filtering algorithms.

% Intractable likelihoods
However for some models of interest, it is not possible to make use of \textsc{smc} due to that $g_{\theta}( y_{t  } | x_t)$ lacks an analytical closed-form expression, is defined recursively or is computationally prohibitive to evaluate. We refer to this class of models as \textit{\textsc{ssm}s with intractable likelihoods}. One example is when the $\alpha$-stable distribution \citep{Nolan2003} is used as $g_{\theta}$ in \eqref{eq:ssmDef} to model heavy-tailed noise in the observations. This type of modelling has recently been advocated by \cite{StoyanovRachevaRachevFabozzi2010} among others to capture the behaviour of financial indices and stock prices, which often exhibit so-called jumps.

% SMC-ABC
One approach to obtain a biased estimate of the intractable likelihood is to make use of approximate Bayesian computations (\textsc{abc}; \citealp{MarinPudloRobertRyder2012}) in combination with \textsc{smc} \citep{JasraSinghMartinMcCoy2012}. The parameters $\theta$ can then be estimated using standard inference algorithms. However, the estimator obtained by \textsc{smc}-\textsc{abc} often suffers from a large variance and is computationally expensive to evaluate. This usually results in long run-times (days) of the complete inference algorithm, which is prohibitive in practical applications.

% Our approach
In this paper, we propose a computationally efficient algorithm for Bayes- ian inference in \textsc{ssm}s with intractable likelihoods. The proposed algorithm is referred to as \textsc{gsa} and is a combination of Gaussian process optimisation (\textsc{gpo}; \citealp{BrochuCoraDeFreitas2010}) and \textsc{smc}-\textsc{abc}. The aim of \textsc{gsa} is to construct a Laplace approximation to approximate \eqref{eq:ParameterPosteriorDef}. The efficiency of the proposed algorithm stems from that \textsc{gpo} requires substantially less posterior evaluations and is more robust to noise compared with other optimisation algorithms. This is mainly due to that \textsc{gpo} operates by constructing a \textit{surrogate function} that mimics \eqref{eq:ParameterPosteriorDef} in analogue with \cite{Wood2010}. The resulting surrogate is smooth and computationally cheap to evaluate, which enables the use of standard optimisation methods to extract a Laplace approximation of the surrogate mimicking the true posterior.

%Results and advantages
The main contribution of this paper is to introduce, develop and numerically study the \textsc{gsa} algorithm. We compare the proposed algorithm to particle Metropolis-Hastings (\textsc{pmh}; \citealp{AndrieuDoucetHolenstein2010,DahlinSchon2015}) and \textsc{spsa} \citep{Spall1998,EhrlichJasraKantas2015} for inference in \textsc{ssm}s using both synthetic and real-world data. In Bayesian inference, \textsc{pmh} is seen as the gold standard for posterior approximations and \textsc{spsa} is known as an efficient and scalable gradient-free optimisation algorithm. The numerical comparisons indicate that \textsc{gsa} can: (i) provide good posterior approximations, (ii) reduce the computational time by between one and two orders of magnitude compared with \textsc{pmh}, (iii) exhibit good robustness to the \textsc{abc} approximation and noise in the estimates. Furthermore, we demonstrate how to make use of the proposed algorithm in estimating the risk in financial portfolios.

% Related work
Related work to the proposed algorithm is presented in e.g.\ \citet{DahlinLindsten2014}, \cite{GutmannCorander2015} and \cite{MeedsWelling2014}. In the first two works, the authors aim to obtain a maximum likelihood estimate and \textsc{map} estimate using \textsc{gpo}, respectively. In the present work, we would like to approximate the entire posterior and not only the parameter that maximises the value of the likelihood or posterior. Moreover, compared with \cite{MeedsWelling2014}, the uncertainty encoded into the surrogate function is utilized to determine the next point in which to sample the log-posterior. 

We continue with Section~\ref{sec:overview}, where an overview of the proposed algorithm and its components are presented. Sections~\ref{sec:smc} and \ref{sec:gpo} discuss the details of these components and the resulting algorithm is presented in Section~\ref{sec:algorithm}. We conclude the paper with an extensive numerical evaluation in Section~\ref{sec:results}, and some remarks and future work in Section~\ref{sec:conclusions}.

%%%%%%%%%%%%%%%%%%%%%%%%%%%%%%%%%%%%%%%%%%%%%%%%%%%%%%%%%%%%%%%%%%%%%%%%%%%%%%%%%%%%%%%%%%%%%%%%%%%%%%%
%%%%%%%%%%%%%%%%%%%%%%%%%%%%%%%%%%%%%%%%%%%%%%%%%%%%%%%%%%%%%%%%%%%%%%%%%%%%%%%%%%%%%%%%%%%%%%%%%%%%%%%
%%%%%%%%%%%%%%%%%%%%%%%%%%%%%%%%%%%%%%%%%%%%%%%%%%%%%%%%%%%%%%%%%%%%%%%%%%%%%%%%%%%%%%%%%%%%%%%%%%%%%%%
\section{An intuitive overview of GSA}
\label{sec:overview}
% MAP estimator and why we cannot use it directly
Our aim is to find \textit{Laplace approximation} of the parameter posterior \eqref{eq:ParameterPosteriorDef},
\begin{align}
\widehat{p}(\theta|y_{1:T}) 
&= 
\mathcal{N}
\bigg(
\theta;
\widehat{\theta}_{\text{MAP}} 
,
\Big[ \underbrace{- \nabla^2 \log p(\theta | y_{1:T}) \Big|_{\theta = \widehat{\theta}_{\text{MAP}} }}_{\triangleq \mathcal{J}(\widehat{\theta}_{\text{MAP}} )} \Big]^{-1} \bigg),
\label{eq:laplaceapproximation}
\end{align}
where $\mathcal{N}(\theta; \mu, \Sigma)$ denotes the Gaussian distribution with mean $\mu$ and covariance matrix $\Sigma$. Here, $\mathcal{J}(\widehat{\theta}_{\text{MAP}} )$ denotes the estimate of the Hessian of the log-posterior evaluated at the posterior mode,
\begin{align}
	\widehat{\theta}_{\text{MAP}}
	= 
	\argmax_{\theta \in \Theta}
	\log p(\theta | y_{1:T}),
	\label{eq:MAPestimationProblem}
\end{align}
where $y_{1:T}$ denotes the recorded observations. This can be seen as a Gaussian approximation around the mode of the posterior motivated by the Bernstein-von Mises theorem, which states that the posterior concentrates to a Gaussian distribution centred at the true parameters with the inverse expected information matrix as its covariance when $T \rightarrow \infty$. Note that, even if this is an asymptotic results it can provide reasonable approximations using a finite number of samples as discussed by \citet{PanovSpokoiny2015}.

\begin{figure}[t]
	\centering
	\includegraphics[width=\textwidth]{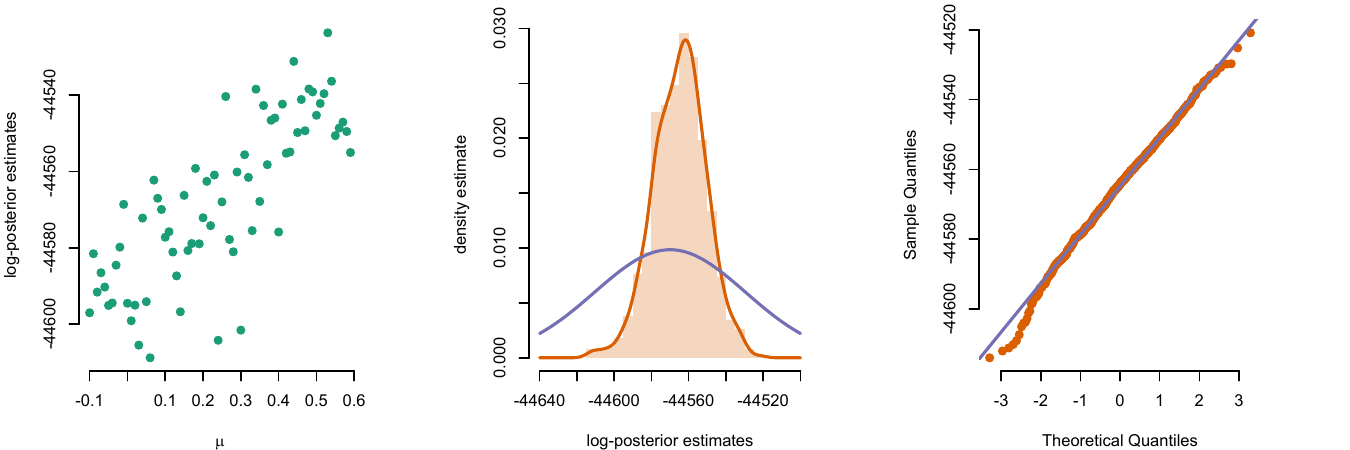}
	\caption{Samples from the log-posterior (left) of \eqref{eq:aSVmodel4parameters} with respect to $\mu$. The distribution (center) and \textsc{qq}-plot(right) of $1,000$ log-posterior estimates of \eqref{eq:aSVmodel4parameters}. The purple lines indicate the best Gaussian approximation. }
	\label{fig:likelihoodestimator}
\end{figure}

We encounter two main problems when constructing the Laplace approximation: (i) the optimisation problem in \eqref{eq:MAPestimationProblem} is difficult to solve efficiently and (ii) $\mathcal{J}(\widehat{\theta}_{\text{MAP}} )$ is typically difficult to estimate with good accuracy. The first problem is due to the high variance in and computational cost of the posterior estimates. An example of this problem is presented in the left part of Figure~\ref{fig:likelihoodestimator}. Here, we present the log-posterior estimates obtained by \textsc{smc}-\textsc{abc} over a grid of $\mu$ in \eqref{eq:aSVmodel4parameters}. The high variance typically results in a slow convergence of parameter inference algorithms such as \textsc{spsa} and \textsc{pmh}. It is also difficult to speed up the convergence by using gradient information as this requires running computationally expensive particle smoothing algorithms.

% 	\item Overview of the method (three iterative steps, suitable when the objective is noisy and costly to evaluate, global optimization, optimise over the surrogate instead and construct the surrogate cleverly by balancing exploration and exploitation).
Instead, we propose to circumvent these problems by optimising a smooth surrogate function that mimics the log-posterior distribution similar to \cite{Wood2010}. We can then obtain a Laplace approximation using standard optimisation methods as the surrogate function is smooth and cheap to evaluate. 

The surrogate function is obtained by \textsc{gpo} algorithm, which is a specific instance of Bayesian optimisation \citep{BrochuCoraDeFreitas2010}. The surrogate function is sequentially updated using samples from the log-posterior. In this paper, we make use of the predictive distribution of a Gaussian process (\textsc{gp}) as the surrogate function. Hence, we can encode certain prior knowledge regarding the smoothness of the log-posterior into the \textsc{gp} prior, which reduces the number of samples required to explore the log-posterior.

The resulting \textsc{gsa} algorithm iterates three steps. At the $k$th iteration:
\begin{itemize}
	\item[(i)] compute an approximation of the log-posterior at the parameter $\theta_k$ denoted $\xi_k = \log \widehat{p}(\theta_k|y_{1:T})$ using \textsc{smc}-\textsc{abc}.
	\item[(ii)] construct a surrogate function by a \textsc{gp} predictive posterior
using the observed data $\{\bm{\theta}_k, \bm{\xi}_k\} = \{\theta_j,\xi_j\}_{j=1}^k$.
	\item[(iii)] evaluate the \textit{acquisition rule} to determine $\theta_{k+1}$.
\end{itemize}

The \textsc{gpo} algorithm then returns an estimate of the posterior and its uncertainty. There are two major advantages of \textsc{gp}s for estimating the posterior density compared with e.g., using splines. Firstly, the uncertainty quantification can be used to develop so-called acquisition rules that explore areas with large uncertainty and exploits the information about the possible location of the mode of the posterior. This typically results in a rapid convergence of the algorithm, which limits the number of posterior samples required and hence also decreases the computational cost. Secondly, the \textsc{gp} can handle noisy function evaluations in a natural manner.

%%%%%%%%%%%%%%%%%%%%%%%%%%%%%%%%%%%%%%%%%%%%%%%%%%%%%%%%%%%%%%%%%%%%%%%%%%%%%%%%%%%%%%%%%%%%%%%%%%%%%%%
%%%%%%%%%%%%%%%%%%%%%%%%%%%%%%%%%%%%%%%%%%%%%%%%%%%%%%%%%%%%%%%%%%%%%%%%%%%%%%%%%%%%%%%%%%%%%%%%%%%%%%%
%%%%%%%%%%%%%%%%%%%%%%%%%%%%%%%%%%%%%%%%%%%%%%%%%%%%%%%%%%%%%%%%%%%%%%%%%%%%%%%%%%%%%%%%%%%%%%%%%%%%%%%
\section{Estimating $\log p(\theta|y_{1:T})$}
\label{sec:smc}
In this section, we discuss how to estimate the log-posterior that is required for carrying out Step~(i) of \textsc{gsa}. This is done by making use of the (bootstrap) particle filter from which we can obtain an estimate of the \textit{marginal filtering distribution} $p_{\theta}(x_{t}|y_{1:t})$ by
\begin{align}
	p^N_{\theta}( \dn x_{t}|y_{1:t})
	= 
	\sum_{i=1}^N
	w_{t}^{(i)}
	\delta_{x_{t}^{(i)}}(\dn x_{t}),
	\label{eq:smc:empFilteringDist}
\end{align}
where $x_{t}^{(i)}$ and $w_{t}^{(i)}$ denotes the particle $i$ at time $t$ and its normalised weight, respectively. Here, $\delta_x$ denotes the Dirac measure placed at $x$. As we shall see, the particle system $\{w_t^{(i)},x_t^{(i)}\}_{i=1}^N$ can also be used to obtain estimates of the log-posterior.

%======================================================================================================
%======================================================================================================
\subsection{Particle filtering with ABC}
\label{sec:smc:pf}
The main problem with applying the particle filter is that it assumes that we can evaluate $g_{\theta}(y_t|x_t)$ point-wise. In the current setting, this is not possible and instead we circumvent the evaluation of  $g_{\theta}(y_t|x_t)$ by using \textsc{abc}. This amounts to augmenting the posterior with an auxiliary variable $\check{y}_{1:T}$, which is data simulated from $g_{\theta}(y_t|x_t)$ for $t=1,\ldots,T$. The fundamental assumption of \textsc{abc} is that data $\check{y}_t$ generated from $g_{\theta}(y_t|x_t)$ should be similar to the observed data $y_t$ if $\theta$ is properly selected. The resulting augmented posterior can be expressed as
\begin{align*}
  	p_{\epsilon}(\theta, \check{y}_{1:T} | y_{1:T} )
  	= 
	\frac{ p(\theta) p_{\theta} \Big( \check{y}_{1:T} \Big)  \rho \Big( \check{y}_{1:T}; y_{1:T}, \epsilon \Big) }
	{ \displaystyle
	\int_{\Theta} \!  p(\theta') p_{\theta'} \Big( \check{y}_{1:T} \Big)  \rho \Big(\check{y}_{1:T}; y_{1:T}, \epsilon \Big) \dd \theta'},
\end{align*}
where, $\rho(\mu,\epsilon)$ denotes some density with mean $\mu$ and tolerance parameter $\epsilon$. This density is used to compute the distance between the simulated observations $\check{y}_t$ and the true observations $y_t$. A common choice is the Gaussian density $\rho(\mu,\epsilon) = \mathcal{N}(\mu,\epsilon^2)$. Finally, we assume that the following marginalisation property holds
\begin{align*}
  	p_{\epsilon}(\theta | y_{1:T} )
  	= 
  	\dint
  	p_{\epsilon}(\theta, \check{y}_{1:T} | y_{1:T} )
  	\dd \check{y}_{1:T},
\end{align*}
when the \textit{tolerance parameter} is \textit{small enough} and where $p_{\epsilon}(\theta| y_{1:T} )$ denotes the posterior of the perturbed model. Hence, we recover a perturbed version of the posterior when $T \rightarrow \infty$ and $\epsilon$ is small enough.

To make use of this in the particle filter, we reformulate the \textsc{ssm} in \eqref{eq:ssmDef} using \textsc{abc} as in \citet{JasraSinghMartinMcCoy2012}. The observed data $y_{1:T}$ is perturbed by
\begin{align}
	\check{y}_t 
	= 
	\psi( y_t ) 
	+ 
	z_t, 
	\quad 
	z_t 
	\sim 
	\rho(0, \epsilon),
	\label{eq:smc:perturbationY}
\end{align}
where $\psi$ denotes some suitable one-to-one transformation. Moreover, we assume that it is possible to simulate from the model using a transformation of random variables. This corresponds to that we can write $\check{y}_t = \tau_{\theta}(\check{x}_{t})$, where $\tau_{\theta}$ denotes a function of $\check{x}_t^{\top}=(x_t^{\top},v_t^{\top})$ and $v_t \sim \nu_{\theta}(v_t|x_t)$ for some probability distribution $\nu_{\theta}$. This is a useful construction as it is often possible to generate samples from complicated distributions but not to evaluate them point-wise. See \ref{app:implementationdetails} for how to select $\tau_{\theta}$ and $\nu_{\theta}$ to generate $\alpha$-stable random variables.

From these two steps, we can rewrite the \textsc{ssm} \eqref{eq:ssmDef} as
\begin{subequations}
\begin{align}
	\check{x}_{t} | \check{x}_{t-1} 
	&\sim 
	\Xi_{\theta}( \check{x}_{t} | \check{x}_{t-1} ) 
	= \nu_{\theta}(v_{t}|x_{t}) f_{\theta}(x_{t}|x_{t-1}), \\
	\check{y}_{t} | \check{x}_{t}
	&\sim 
	h_{\theta,\epsilon}( \check{y}_{t} | \check{x}_{t}) 
	= \rho \Big( \check{y}_t; \psi( \tau_{\theta}( \check{x}_t ) ), \epsilon \Big).
\end{align} 
\label{eq:smc:reformulatedSSM}%
\end{subequations}%
We can now apply a particle filter as in Algorithm~\ref{alg:smc} for this new model, which does not require us to evaluate the $g_{\theta}(y_t|x_t)$ point-wise but only that we can simulate $\check{y}_{t}$. In this paper, we leave the choice of $\epsilon$ to the user, which can be done using e.g.\ pilot runs on simulated data. However, it is possible to adapt $\epsilon$ on-the-fly using the approaches discussed by e.g.\ \cite{DelMoralDoucetJasra2012} or \cite{CalvetCzellar2015}. However in our experience, these methods require a larger value of $N$ than un-adapted \textsc{smc}-\textsc{abc} to provide log-posterior estimates with a reasonable bias. Therefore, we decided to fix $\epsilon$ in this paper to obtain computationally efficient algorithms.

\begin{algorithm}[!t]
\caption{\textsf{Estimate the log-posterior by particle filtering}}
\textsc{Inputs:} $\check{y}_{1:T}$ (perturbed data), \textsc{ssm} \eqref{eq:smc:reformulatedSSM}, $N \in \mathbb{N}$ (no.\ particles), $\rho_{\epsilon}$ (density for \textsc{abc} approximation) with $\epsilon \in \mathbb{R}_+$ (tolerance par.). \\
\textsc{Outputs:} $\log \widehat{p}^N(\theta|\check{y}_{1:T})$ (est.\ of log-posterior). \\
\textsc{Note:} all operations are carried out over $i,j = 1, \ldots, N$.
\algrule[.4pt]
\begin{algorithmic}[1]
	\STATE Sample $\check{x}^{(i)}_0 \sim \mu_{\theta}(x_0) \nu_{\theta}(v_0|x_0)$ and set $w_0^{(i)}=1/N$.
	\FOR{$t=1$ to $T$}
		\STATE Apply systematic resampling to obtain the \textit{ancestor index} $a^{(i)}_t$ from a categorical distribution with $\mathbb{P} \Big( a^{(i)}_t = j \Big) =  w^{(j)}_{t-1}.$
		\STATE Propagate particles using $\check{x}^{(i)}_t \sim \Xi_{\theta} \Big( \check{x}_{t} | \check{x}_{t-1}^{ a_t^{(i)} } \Big)$ and set $\check{x}_{0:t}^{(i)} = \Big\{ \check{x}_{0:t-1}^{a_t^{(i)}}, \check{x}_{t}^{(i)} \Big\}$.
		\STATE Compute particle weights by $W_t^{(i)} = h_{\theta,\epsilon} \Big( \check{y}_{t} | \check{x}_{t}^{(i)} \Big)$ and			$
			w^{(i)}_t = W_t^{(i)} \left[ \sum_{j=1}^N W_t^{(j)} \right]^{-1}.
		$
	\ENDFOR
	\STATE Compute $\log \widehat{p}^N(\theta|\check{y}_{1:T})$ by \eqref{eq:smc:loglikeEst}.
\end{algorithmic}
\label{alg:smc}
\end{algorithm}

%======================================================================================================
%======================================================================================================
\subsection{The estimator and its statistical properties}
\label{sec:smc:loglikelihoodest}
An estimator for the log-posterior of the perturbed model \eqref{eq:smc:reformulatedSSM} is given by
\begin{align}
	\log \widehat{p}^N( \theta | y_{1:T} )
	&=
	\log \widehat{p}^N_{\theta}( y_{1:T} )
	+
	\log p(\theta) \nonumber \\
	&=
	\sum_{t=1}^T \log \left\{ \sum_{i=1}^N W_{t}^{(i)} \right\}
	- T \log N
	+
	\log p(\theta),
	\label{eq:smc:loglikeEst}
\end{align}
which makes use of the unnormalised weights generated by Algorithm~\ref{alg:smc}. It is known from \citet{PittSilvaGiordaniKohn2012} that the estimator \eqref{eq:smc:loglikeEst} is biased for a finite number of particles but it is consistent and asymptotically Gaussian. Specifically, we have that the error in the log-posterior estimate fulfils a \textsc{clt} given by 
\begin{align}
  \sqrt{N} 
  \bigg[ 
  \log p( \theta | y_{1:T} ) 
  - 
  \log \widehat{p}^N( \theta | y_{1:T} )
  + 
  \frac{\gamma^2(\theta)}{2N} \bigg] 
  \stackrel{d}{\longrightarrow} 
  \mathcal{N} \Big( 0,\gamma^2(\theta) \Big),
  \label{eq:smc:loglikeEstCLT}
\end{align}
when $N \rightarrow \infty$ and for some unknown variance $\gamma(\theta)$. As a result, we have an expression for the bias of the estimator given by $-\gamma^2(\theta)/2N$ for a finite number of particles. However, we see that the error is approximately Gaussian for this type of model in the finite sample case by the experimental data presented in the center and right parts of Figure~\ref{fig:likelihoodestimator}. 

The log-posterior estimator in \eqref{eq:smc:loglikeEst} is consistent with respect to the perturbed model \eqref{eq:smc:reformulatedSSM} but \textit{not} the true model \eqref{eq:ssmDef}. \citet{DeanSingh2011} show that the perturbation results in a bias in the parameter estimates (w.r.t.\ the unperturbed model) that decrease proportional to $\mathcal{O}(\epsilon^2)$ under some regularity assumptions. Furthermore, the asymptotic Gaussianity of the estimator and a Bernstein-von Mises-type theorem holds for a small enough $\epsilon$. This is an important fact to motivate the Laplace approximation as discussed in Section~\ref{sec:overview} and we investigate these properties empirically in Section~\ref{sec:results:Gaussian:bias}.

%%%%%%%%%%%%%%%%%%%%%%%%%%%%%%%%%%%%%%%%%%%%%%%%%%%%%%%%%%%%%%%%%%%%%%%%%%%%%%%%%%%%%%%%%%%%%%%%%%%%%%%
%%%%%%%%%%%%%%%%%%%%%%%%%%%%%%%%%%%%%%%%%%%%%%%%%%%%%%%%%%%%%%%%%%%%%%%%%%%%%%%%%%%%%%%%%%%%%%%%%%%%%%%
%%%%%%%%%%%%%%%%%%%%%%%%%%%%%%%%%%%%%%%%%%%%%%%%%%%%%%%%%%%%%%%%%%%%%%%%%%%%%%%%%%%%%%%%%%%%%%%%%%%%%%%
\section{Constructing the surrogate of $\log p(\theta|y_{1:T})$}
\label{sec:gpo}
In this section, we briefly discuss Steps~(ii) and (iii) of \textsc{gsa}, where we construct a surrogate function to mimic the log-posterior. The interested reader is referred to \citet{BrochuCoraDeFreitas2010} for more details.

%======================================================================================================
%======================================================================================================
\subsection{Gaussian process prior}
\label{sec:gpo:predictiveposterir}
\textsc{gp}s \citep{RasmussenWilliams2006} are an instance of so-called \textit{Bayesian non-parametric models} and can be interpreted as a generalisation of the multivariate Gaussian distribution to an infinite dimensional setting. A realisation drawn from a \textsc{gp} can therefore be seen as an infinite vector of real values (a function over the real space $\mathbb{R}^p$). To construct the surrogate function, we assume a priori that the log-posterior is distributed according to
\begin{align}
	\log p(\theta | y_{1:T}) \sim \mathcal{GP} \big( m(\theta),\kappa(\theta,\theta') \big).
	\label{eq:GPprior}
\end{align}
Note that this does \textit{not} correspond to an assumption that the log-posterior of the parameters in the \textsc{ssm} is Gaussian. Here, $\mathcal{GP}(m,\kappa)$ denotes a \textsc{gp} with mean function $m$ and covariance function $\kappa$ defined by
\begin{align*}
	m(\theta) &= \mathbb{E} \Big[ \log p(\theta | y_{1:T}) \Big], \\
	\kappa(\theta,\theta') &= \mathbb{E} \Big[  \Big( \log p(\theta | y_{1:T}) - m(\theta) \Big) \Big( \log p(\theta' | y_{1:T}) - m(\theta') \Big) \Big].
\end{align*}
The mean function specifies the expected value of the process and the covariance function specifies the dependence between any pair of points on the log-posterior function. The covariance function depends on a set of hyperparameters, such as the \textit{length scale} that controls the dependence, see \cite{RasmussenWilliams2006} for details. From the \textsc{clt} in \eqref{eq:smc:loglikeEstCLT} and Figure~\ref{fig:likelihoodestimator}, we know that the error in the log-posterior is approximately Gaussian,
\begin{align}
	\xi_k = 
	\log \widehat{p}^N(\theta_k|y_{1:T}) 
	\approx
	\log p(\theta_k|y_{1:T}) + \sigma_{\xi} z_k, 
	\qquad 
	z_k \sim \mathcal{N}(0,1),
	\label{eq:GPregModel}
\end{align}
where $\sigma^2_{\xi}$ denotes some unknown variance estimated in a later stage of the algorithm. Consequently, we have that the \textit{predictive posterior} for any test point $\theta_{\star} \in \Theta$ is given by
\begin{subequations}
\begin{align}
	\log p(\theta_{\star}|y_{1:T}) | \mathcal{D}_k 
	&\sim 
	\mathcal{GP} \Big( 
	\mu(\theta_{\star}|\mathcal{D}_k),
	\sigma^2(\theta_{\star}|\mathcal{D}_k) + \sigma^2_{\xi}
	\Big),
	\\
	\mu(\theta_{\star}|\mathcal{D}_k)
	&=
	m(\theta_{\star}) \nonumber \\ &+ \kappa(\theta_{\star},\bm{\theta}_k) 
	\Big[ \kappa(\bm{\theta}_k,\bm{\theta}_k) + \sigma^2_{\xi} \mathbf{I}_{k \times k} \Big]^{-1} 
	\Big\{ \bm{\xi}_k - m(\theta_{\star}) \Big\}, \\
	\sigma^2(\theta_{\star}|\mathcal{D}_k)
	&=
	\kappa(\theta_{\star},\theta_{\star}) \nonumber \\ &- \kappa(\theta_{\star},\bm{\theta}_k) 	
	\Big[ \kappa(\bm{\theta}_k,\bm{\theta}_k) + \sigma^2_{\xi} \mathbf{I}_{k \times k} \Big]^{-1}
	\kappa(\bm{\theta}_k,\theta_{\star}),
\end{align}%
\label{eq:gpo:GPsurrogate}%
\end{subequations}%
where we have introduced the notation $\mathcal{D}_k = \{\bm{\theta}_k, \bm{\xi}_k\}$ for the information available at iteration $k$. The surrogate function of the log-posterior is then given by \eqref{eq:gpo:GPsurrogate}. The major cost in computing the predictive posterior is incurred by the matrix inversion which is proportional to $\mathcal{O}(k^3)$. Hence, sparse formulations of the \textsc{gp} can be useful to decrease the computation cost for large $K$, see \cite{RasmussenWilliams2006}.

%======================================================================================================
%======================================================================================================
\subsection{Acquisition function}
\label{sec:gpo:aqfunc}
The surrogate function given by the \textsc{gp} predictive distribution gives us the estimate of the log-posterior and its uncertainty. As previously mentioned, this is useful information for creating an acquisition rule $\mathsf{AQ}(\theta_{\star}|\mathcal{D}_k)$ to balance exploration and exploitation. We can then determine the next point in which to sample the log-posterior by
\begin{align*}
		\theta_{k+1} 
		= 
		\argmax_{\theta_{\star} \in \Theta_{\text{GPO}}} \mathsf{AQ}(\theta_{\star}|\mathcal{D}_k),
\end{align*}
where $\Theta_{\text{GPO}}$ denotes a \textit{search space} defined by the user. In this paper, we make use of the \textit{expected improvement} (\textsc{ei}) due to its general good performance in numerical evaluations \citep{Lizotte2008}. To derive the \textsc{ei} rule, consider the \textit{predicted improvement} defined as
\begin{align}
	\mathsf{PI}(\theta_{\star})
	&= 
	\max \Big\{ 
	0, \log p(\theta_{\star}|y_{1:T}) - \mu_{\max} - \zeta
	\Big\}, \qquad \forall \theta_{\star} \in \Theta_{\text{GPO}},
	\label{eq:gpo:predictedImprovement}
\end{align}
where $\mu_{\max}$ denotes the maximum value of $\mu(\theta)$ for the sampled points $\theta \in \bm{\theta}_k$. Here, we introduce $\zeta$ as a parameter that controls the exploitation/exploration behaviour as in \citet{Lizotte2008}. Hence for $\zeta=0$, we have that $\mathsf{PI}(\theta_{\star})$ is the difference between the posterior and the maximum value it assumes in the set of sampled points. Therefore, it is positive for points where the log-posterior is larger than the current peak and zero for all other points. The \textsc{ei} rule is obtained by computing the expected value of \eqref{eq:gpo:predictedImprovement} with respect to \eqref{eq:gpo:GPsurrogate}. This results in the acquisition rule given by
\begin{align}
	\theta_{k+1} &= 
	\left\{
	\argmax_{\theta_{\star} \in \Theta_{\text{GPO}}} \,
	\sigma(\theta_{\star}|\mathcal{D}_k)
	\Big[ Z(\theta_{\star}) \Phi \big( Z(\theta_{\star}) \big) + \phi \big( Z(\theta_{\star}) \big) \Big]
	\right\}
	 + \check{z}_k, 
	\label{eq:gpo:aqoptimisation}
	\\
	Z(\theta_{\star}) 
	&=
	\sigma^{-1}(\theta_{\star}|\mathcal{D}_k)
	\Big[
	\mu(\theta_{\star}|\mathcal{D}_k) - \mu_{\max} - \zeta
	\Big],
	\nonumber
\end{align}
where $\phi$ and $\Phi$ denotes the density and distribution function of the standard Gaussian distribution, respectively. Here, we \textit{jitter} to the solution of the optimisation problem by adding Gaussian noise $\check{z}_k \sim \mathcal{N}(0,\Sigma)$ with covariance $\Sigma$. In practice, this improves the exploration and increases the accuracy of the obtained parameter estimates. Jittering is also advocated by \citet{Bull2011} and \citet{GutmannCorander2015} to increase the convergence rate of \textsc{gpo}.

The optimisation in \eqref{eq:gpo:aqoptimisation} is possibly non-convex but it is cheap to evaluate the objective function as it only amounts to evaluating the \textsc{gp} predictive posterior in one point. We make use of the the gradient-free dividing rectangles (\textsc{direct}; \citealp{Jones1993}) to solve \eqref{eq:gpo:aqoptimisation} over $\Theta_{\text{GPO}}$, which is determined from the support of the prior distribution $p(\theta)$.

%%%%%%%%%%%%%%%%%%%%%%%%%%%%%%%%%%%%%%%%%%%%%%%%%%%%%%%%%%%%%%%%%%%%%%%%%%%%%%%%%%%%%%%%%%%%%%%%%%%%%%%
%%%%%%%%%%%%%%%%%%%%%%%%%%%%%%%%%%%%%%%%%%%%%%%%%%%%%%%%%%%%%%%%%%%%%%%%%%%%%%%%%%%%%%%%%%%%%%%%%%%%%%%
%%%%%%%%%%%%%%%%%%%%%%%%%%%%%%%%%%%%%%%%%%%%%%%%%%%%%%%%%%%%%%%%%%%%%%%%%%%%%%%%%%%%%%%%%%%%%%%%%%%%%%%
\section{The GSA algorithm}
\label{sec:algorithm}
% Combine the previous sections into Algorithm~2
\textsc{gsa} (Algorithm~\ref{alg:GPOABC}) is obtained by combining \textsc{smc}-\textsc{abc} (Algorithm~\ref{alg:smc}) to approximate the log-posterior point-wise and \textsc{gpo} to create a surrogate function that mimics the log-posterior around its mode. In this section, we discuss some user choices and convergence results for the algorithm. See \ref{app:implementationdetails} for the details of the implementation employed in this paper.

\subsection{Initialisation and convergence criteria}
% Discuss initalisation of the GSA algorithm
We initialise Algorithm~\ref{alg:GPOABC} at Line~1 to find some suitable hyperparameters for the \textsc{gp} prior. The hyperparameters are estimated using $L$ initial samples from the log-posterior obtained by Latin hypercube sampling. We then execute Algorithm~\ref{alg:smc} for each of the sampled parameters $\{\theta_1^{\star},\theta_2^{\star},\ldots,\theta_L^{\star}\}$ to obtain $\mathcal{D}^{\star}_L$ by the analogue of Line~4 in Algorithm~\ref{alg:GPOABC}. After the initialisation of the algorithm, we can update the hyperparameters with Line~5 at every iteration or at some pre-defined interval. Estimating the hyperparameters is computationally costly and it is therefore recommended to re-estimate them only at some fixed interval. The algorithm is usually executed for some pre-defined number of iterations $K$ or until the \textsc{ei} is smaller than some $\Delta \mathsf{EI} > 0$, i.e.\ until $k$ satisfies $\mathsf{EI}(\theta_k|\mathcal{D}) < \Delta \mathsf{EI}$. 

\begin{algorithm}[!t]
\caption{Find a Laplace approximation of $\log p(\theta|y_{1:T})$ using \textsf{GSA}}
\textsc{Inputs:} Algorithm \ref{alg:smc}, $p(\theta)$ (parameter prior), $m(\theta)$ (mean function), $\kappa(\theta,\theta')$ (covariance function), $\theta_1$ (initial parameter), $\Sigma$ (jittering covariance) and $\Theta_{\text{GPO}}$ (optimisation bounds). \\
\textsc{Output:} $\widehat{\theta}_{\text{MAP}}$ (est.\ of the parameter) and $\widehat{\mathcal{J}}(\widehat{\theta}_{\text{MAP}})$ (est.\ of posterior covariance).
\algrule[.4pt]
\begin{algorithmic}[1]
	\STATE Estimate the hyperparameters of the \textsc{gp} prior by using some initial data $\mathcal{D}^{\star}_L$.
	\STATE Initialise the parameter to $\theta_1$ and set $k=1$.
	\WHILE{\textit{convergence criteria is not satisfied}}
		\STATE Estimate $\xi_k = \log \widehat{p}(\theta_k|y_{1:T})$ by Algorithm \ref{alg:smc} and set $\mathcal{D}_k = \{\mathcal{D}^{\star}_L,\bm{\theta}_k,\bm{\xi}_k\}$.
		\STATE (\textit{if required}) Update the hyperparameters of the \textsc{gp} prior using $\mathcal{D}_k$.		
		\STATE Construct the \textsc{gp} surrogate $\log p(\theta_{\star}|y_{1:T}) | \mathcal{D}_k$ using \eqref{eq:gpo:GPsurrogate}.
		\STATE Compute $\mu_{\max} = \argmax_{\theta \in \bm{\theta}_k} \mu(\theta | \mathcal{D}_k)$.
		\STATE Compute $\theta_{k+1}$ by \eqref{eq:gpo:aqoptimisation} using optimisation over $\Theta_{\text{GPO}}$.
		\STATE Set $k = k +1$.
	\ENDWHILE
	\STATE Compute the \textsc{map} estimate $\widehat{\theta}$ by optimising $\mu(\theta|\mathcal{D}_{k})$ using optimisation over $\Theta_{\text{GPO}}$.
	\STATE Extract the Hessian estimate $\mathcal{J}(\widehat{\theta}_{\text{MAP}} )$ using e.g.\ finite-differences on $\mu(\theta|\mathcal{D}_k)$.
\end{algorithmic}
\label{alg:GPOABC}
\end{algorithm}

\subsection{Extracting the Laplace approximation}
% Discuss how to extract the Laplace approximation
From Algorithm~\ref{alg:GPOABC}, we obtain the predictive posterior mean function $\mu(\theta_{\star}|\mathcal{D})$, which hopefully is an accurate surrogate for the log-posterior. We then proceed to extract the \textsc{map} estimate $\widehat{\theta}_{\text{MAP}}$ defined by \eqref{eq:MAPestimationProblem}. As the surrogate is smooth and cheap to evaluate, we can carry out the optimisation using standard methods such as the \textsc{direct} algorithm to find the mode. The estimate of the Hessian of the log-posterior $\mathcal{J}(\widehat{\theta}_{\text{MAP}} )$ can be computed by a finite-difference scheme, by analytically computing the Hessian of $\mu(\theta_{\star}|\mathcal{D})$ when possible or by using a quasi-Newton algorithm to solve \eqref{eq:MAPestimationProblem}.

\subsection{Convergence properties}
% Discuss theoretical properties
There are only a limited number of results regarding the convergence properties of the \textsc{gpo} algorithm in the literature. Most of the properties have been studied numerically by benchmarking the \textsc{gpo} algorithm against alternatives on a large number of optimisation problems. However, some theoretical results are discussed by \citet{Bull2011} and \citet{VazquezBect2010}. They conclude that \textsc{gpo} using the \textsc{ei} rule samples the log-posterior densely if it is continuous with respect to the \textsc{gp} prior. Also, \textsc{gpo} achieves an \textit{optimal} convergence rate of the order $\mathcal{O} \big( (K \log K )^{-5/p} ( \log K )^{1/2} \big)$ for the Mat\'{e}rn 5/2 covariance function, where $K$ and $p$ denote the number of samples and parameters to infer, respectively.

%%%%%%%%%%%%%%%%%%%%%%%%%%%%%%%%%%%%%%%%%%%%%%%%%%%%%%%%%%%%%%%%%%%%%%%%%%%%%%%%%%%%%%%%%%%%%%%%%%%%%%%
%%%%%%%%%%%%%%%%%%%%%%%%%%%%%%%%%%%%%%%%%%%%%%%%%%%%%%%%%%%%%%%%%%%%%%%%%%%%%%%%%%%%%%%%%%%%%%%%%%%%%%%
%%%%%%%%%%%%%%%%%%%%%%%%%%%%%%%%%%%%%%%%%%%%%%%%%%%%%%%%%%%%%%%%%%%%%%%%%%%%%%%%%%%%%%%%%%%%%%%%%%%%%%%
\section{Numerical illustrations and applications}
\label{sec:results}
In this section, we provide four illustrations of the properties and advantages of the proposed algorithm. The implementation details are collected in \ref{app:implementationdetails} and the source code with data is available for download at GitHub: \url{https://github.com/compops/gpo-smc-abc/}.

%======================================================================================================
%======================================================================================================
\subsection{Stochastic volatility with Gaussian log-returns}
\label{sec:results:Gaussian}
\label{sec:results:Gaussian:bias}
\label{sec:results:Gaussian:computationalcost}
Consider the stochastic volatility model with Gaussian log-returns (\textsc{gsv}),
\begin{subequations}
\begin{align}
	x_0 &\sim \mathcal{N} \left( x_0; \mu, \frac{\sigma^2_v}{ \big( 1-\phi^2 \big) } \right), 
	\\
	x_{t+1} &\sim \mathcal{N} \Big( x_{t+1}; \mu + \phi ( x_t - \mu ), \sigma_v^2 \Big),
	\\
	y_{t}   &\sim \mathcal{N} \Big( y_{t};   0, \exp(x_t) \Big),
\end{align}%
\label{eq:SVmodel3parameters}%
\end{subequations}%
\noindent with parameters $\theta=\{\mu,\phi,\sigma_v\}$. Here, the latent log-volatility is assumed to follow a mean-reverting random walk with mean $\mu \in \mathbb{R}$, persistence $\phi \in (-1,1)$ and standard deviation of the increments $\sigma_v \in \mathbb{R}_+$. We generate a single synthetic data set from this model with $T=500$ observations, parameters $\theta^{\star}=\{0.20,0.96,0.15\}$ and initial state $x_0=0$. 

\begin{figure}[p]
	\centering
	\includegraphics[width=\textwidth]{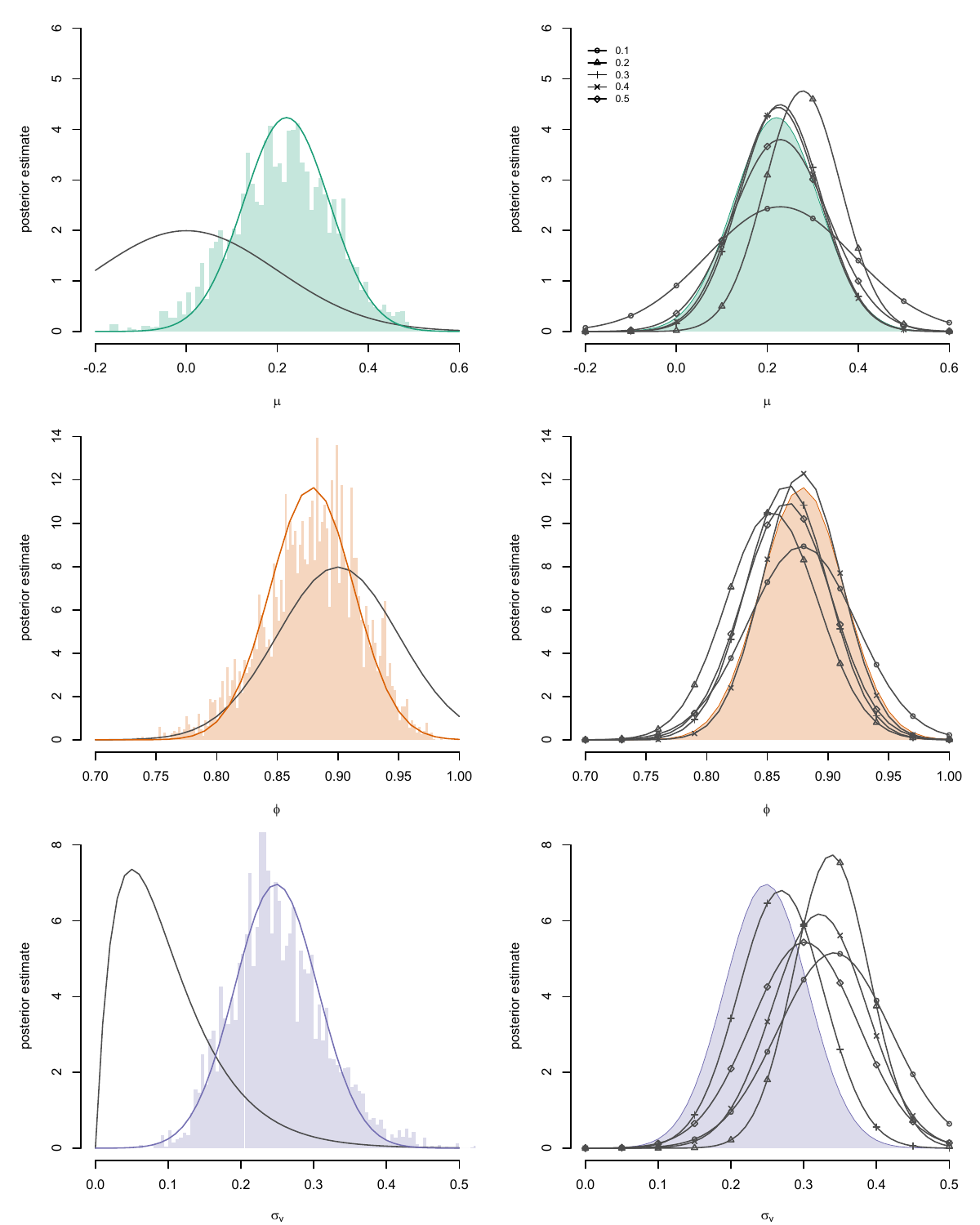}
	\caption{Marginal parameter posteriors for the synthetic data in the \textsc{gsv} model. Left: Solid curves indicate the Laplace approximations of the posterior using \textsc{gs} for $\mu$ (green), $\phi$ (orange) and $\sigma_v$ (purple). The histograms represent the exact posteriors estimated using \textsc{pmh} and the dark grey (left) curves indicate the prior distributions. Right: Laplace approximations (shaded areas) from \textsc{gs} for the three parameters. The grey curves (right) indicate the Laplace approximations obtained by \textsc{gsa} using five different values of the tolerance parameter $\epsilon$ in the \textsc{abc} approximation.}
	\label{fig:example1-posteriors}
\end{figure}

This model is interesting since it allows us to compare the \textsc{gsa} algorithm to an algorithm that makes use of a standard particle filter to estimate the log-posterior. This is possible as we can evaluate $g_{\theta}(y_t|x_t)$ in closed-form for \eqref{eq:SVmodel3parameters}. We refer to this version of the algorithm as \textsc{gs} and it corresponds to the \textsc{gsa} algorithm with tolerance parameter $\epsilon = 0$.

In the left part of Figure~\ref{fig:example1-posteriors}, we present the posterior estimates from \textsc{gs} (solid curve) and \textsc{pmh} (histogram). We begin by observing a good fit of the Laplace approximation from \textsc{gs} to the histogram approximation obtained by \textsc{pmh}; both the location and the spread of the posterior approximations are similar. Using \textsc{gs} results in a speed-up of about $30$ times ($15,000/500 \approx 30$) compared with \textsc{pmh}, where the latter is seen as the gold standard in computational Bayesian inference The main computational cost for both algorithms is incurred by running the particle filter (one run per iteration). The overhead for the proposed algorithm (estimating hyperparameters in the \textsc{gp} and computing the predictive posterior) is negligible compared with the computational cost of running a particle filter.

We now continue with analysing the Laplace approximations obtained by \textsc{gsa}. In the right side of Figure~\ref{fig:example1-posteriors}, we present the posterior approximations obtained by varying the tolerance parameter $\epsilon$ between $0.1$ and $0.5$ to study the bias and robustness of the approximation. Darker shades of grey indicate a larger tolerance parameter. For $\epsilon=0.1$ and $0.2$, we see that the approximations are rather poor with a significant bias and bad fit to the spread. However for the other choices of $\epsilon$, the approximations from \textsc{gsa} converges quickly to be similar to \textsc{gs} (solid curve). From these results, we conclude that there is a bias in the posterior approximation when $\epsilon$ is too small and the approximation tends to grow wider as $\epsilon$ increases. Hence, a bias-variance trade-off is introduced into the posterior approximation depending on $\epsilon$.

\begin{figure}[t]
	\centering
	\includegraphics[width=\textwidth]{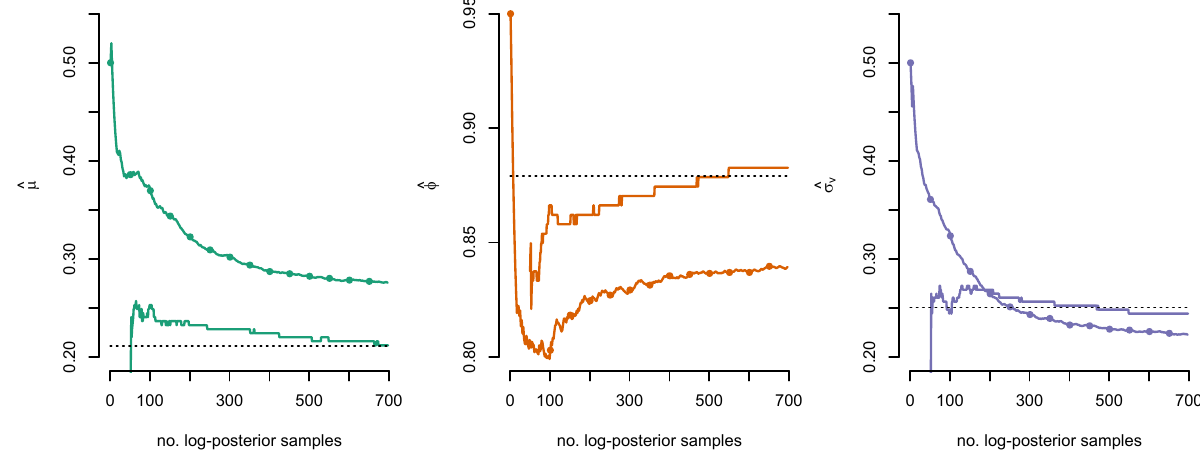}
	\caption{The trace of the \textsc{map} estimate for $\mu$ (green), $\phi$ (orange) and $\sigma_v$ (purple) from \textsc{gs} (solid) and \textsc{spsa} (solid-cicle) as a function of the number of log-posterior samples. The first $L=50$ samples of \textsc{gs} are used to estimate the hyperparameters. Both algorithms are run for a total of $700$ log-posterior samples. Dashed lines indicate the posterior means from \textsc{pmh}.}
	\label{fig:example1-spsa}
\end{figure}

We continue by comparing \textsc{gs} to \textsc{spsa}, where the latter is a gradient-free alternative with good convergence properties and performance in many applications. \textsc{spsa} operates by constructing a finite-difference approximation of the gradient at each iteration after which it takes a step in the gradient direction. Note that \textsc{spsa} requires two log-posterior estimates at each iteration compared with only one sample in the \textsc{gs}. Another possible drawback with \textsc{spsa} is that it only provides the \textsc{map} estimate and no quantification of the posterior uncertainty.

In Figure~\ref{fig:example1-spsa}, we compare the \textsc{map} parameter estimates of two algorithms as a function of the number of log-posterior estimates. The first $L=50$ samples of the log-posterior are used to estimate the hyperparameters of the \textsc{gp} prior. After this initial phase, \textsc{gs} converges quickly to reasonable values of the parameters using about half the number of posterior samples. This results in a speed-up of factor $2$ when using the proposed algorithm compared with \textsc{spsa}.

\begin{figure}[p]
	\centering
	\includegraphics[width=\textwidth]{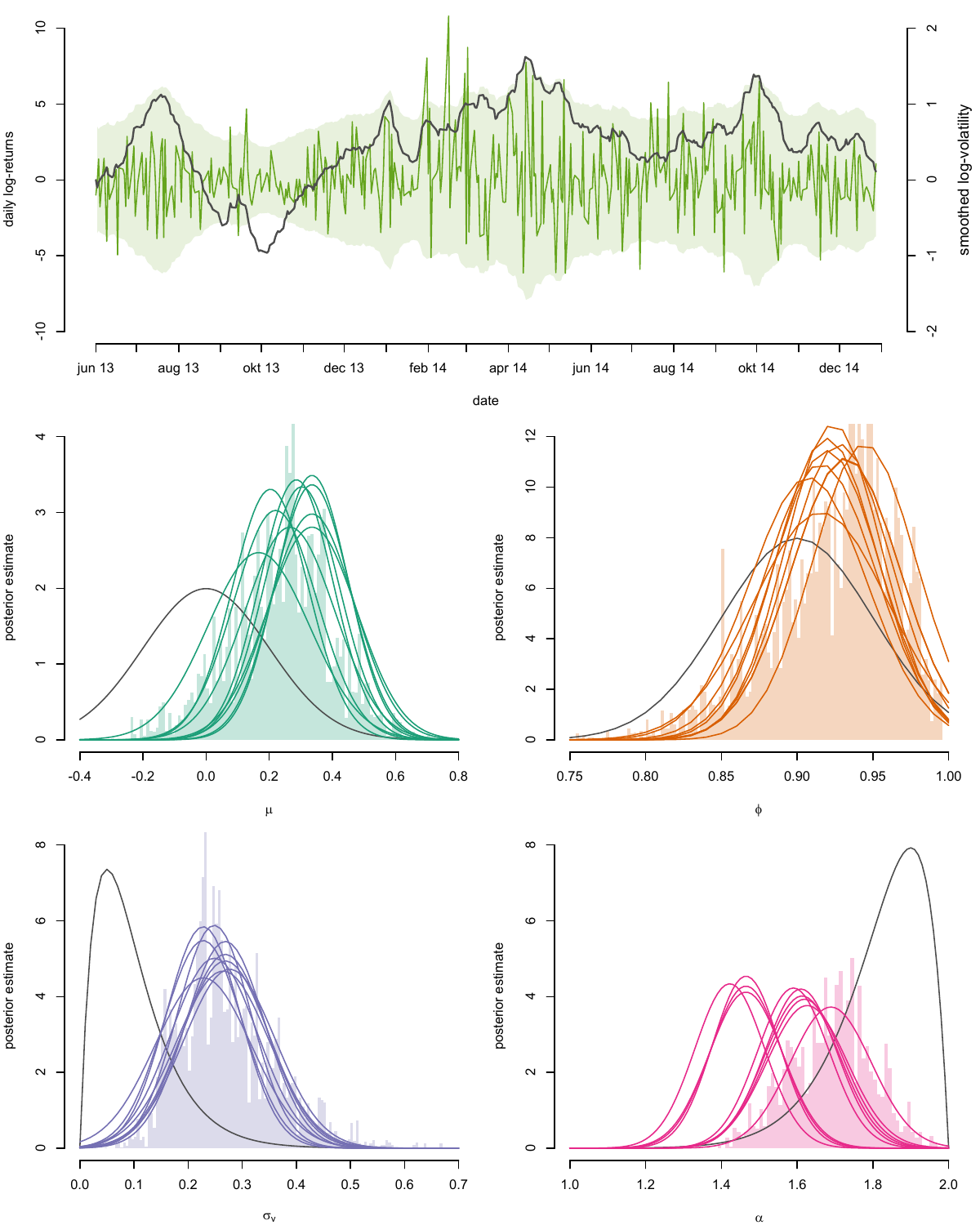}
	\caption{Upper: log-returns (green) of coffee futures and the estimate of the log-volatility (dark grey). The shaded area indicates the $95 \%$ confidence region for the log-returns according to the $\alpha$\textsc{sv} model. Middle and lower: marginal parameter posteriors in the $\alpha$\textsc{sv} model estimated by \textsc{gsa} (solid curves) using $10$ independent runs and \textsc{pmh} (histogram) for $\mu$ (green), $\phi$ (orange), $\sigma_v$ (purple) and $\alpha$ (magenta). Dark grey curves indicate the prior distributions.}
	\label{fig:example2-posteriors}
\end{figure}

%======================================================================================================
%======================================================================================================
\subsection{Stochastic volatility model $\alpha$-stable log-returns}
\label{sec:results:alpha}
In the upper part of Figure~\ref{fig:example2-posteriors}, we present the log-returns of future contracts on coffee during the period between June, 2013 and December, 2014. The data seem to indicate the presence of jumps around the first half of $2014$. This is common in financial data and can be modelled in a number of different ways. To this end, we consider the stochastic volatility model with $\alpha$-stable log-returns ($\alpha$\textsc{sv}) given by

\begin{subequations}
\begin{align}
	x_0 &\sim \mathcal{N} \left( x_0; \mu, \frac{\sigma^2_v}{ \big( 1-\phi^2 \big) } \right), 
	\\
	x_{t+1} &\sim \mathcal{N} \Big( x_{t+1}; \mu + \phi ( x_t - \mu ), \sigma_v^2 \Big),
	\\
	y_{t}   &\sim \mathcal{A} \Big( y_{t};   \alpha, \exp(x_t) \Big),
\end{align}%
\label{eq:aSVmodel4parameters}%
\end{subequations}%

\noindent with parameters $\theta=\{\mu,\phi,\sigma_v,\alpha\}$ and $\mathcal{A}(\alpha,\gamma)$ denoting a zero-mean symmetric $\alpha$-stable distribution with stability parameter $\alpha \in (0,2)$ and scale $\gamma \in \mathbb{R}_+$. The stability parameter determines the tail behaviour of the distribution, see \cite{Nolan2003} for a discussion of the $\alpha$-stable distribution and its properties. The likelihood is in general intractable for this model and therefore approximations such as the particle filtering using \textsc{abc} are required.

In the middle and lower part of Figure~\ref{fig:example2-posteriors}, we compare the posterior approximations obtained with \textsc{gsa} (solid curves) with $10$ independent runs and \textsc{pmh} (histograms). We see that the mixing of \textsc{pmh} is quite poor for this model as the histograms are \textit{peaky}. This is a common problem as Markov chains tends to get stuck if the log-posterior estimates are noisy. However, the posterior estimates overlap and seems to give reasonable parameter values for each run of \textsc{gsa}. The main difference is in the estimate of $\alpha$, which could be the result of the \textsc{abc} approximation or problems with observability. Finally, the estimate the log-volatility (black) seems reasonable when compared to the log-returns.

%%%%%%%%%%%%%%%%%%%%%%%%%%%%%%%%%%%%%%%%%%%%%%%%%%%%%%%%%%%%%%%%%%%%%%%%%%%%%%%%%%%%%%%%%%%%%%%%%%%%%%%
%%%%%%%%%%%%%%%%%%%%%%%%%%%%%%%%%%%%%%%%%%%%%%%%%%%%%%%%%%%%%%%%%%%%%%%%%%%%%%%%%%%%%%%%%%%%%%%%%%%%%%%
%%%%%%%%%%%%%%%%%%%%%%%%%%%%%%%%%%%%%%%%%%%%%%%%%%%%%%%%%%%%%%%%%%%%%%%%%%%%%%%%%%%%%%%%%%%%%%%%%%%%%%%
\subsection{Computing VaR for a portfolio of oil futures}
\label{sec:results:application:oil}
\begin{figure}[p]
	\centering
	\includegraphics[width=0.95\textwidth]{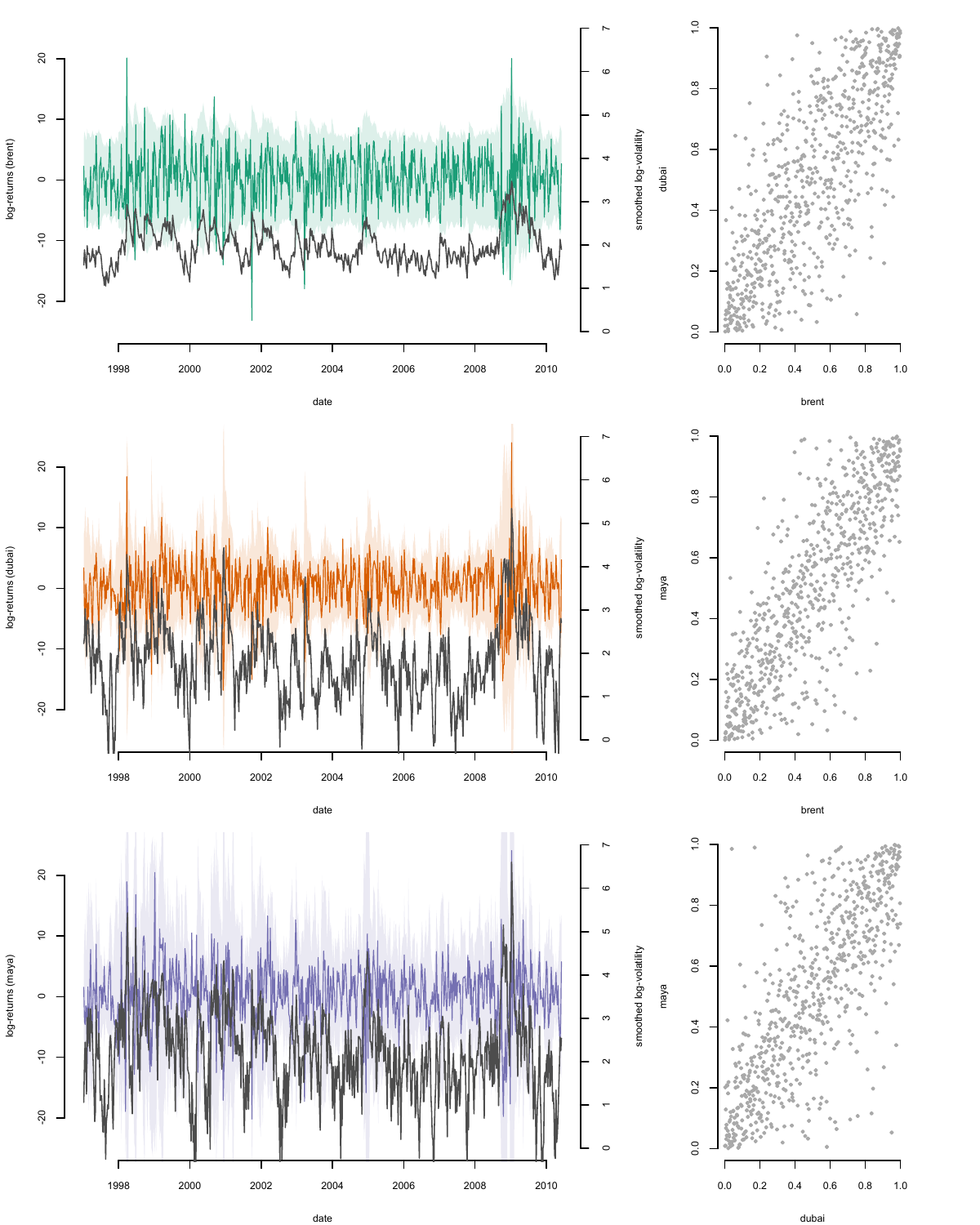}
	\caption{Left: weekly log-returns of Brent (green), Dubai (orange) and Maya (purple) oil between January 10, 1997 and June 4, 2010. The shaded area indicates the $95 \%$ confidence region for the log-returns according to the $\alpha$\textsc{sv} model. The dark grey curves indicate estimates of the log-volatility. Right: the corresponding transformed filtered residuals $\widehat{u}_{t,i}$.}
	\label{fig:example3-oil-copula-data}
\end{figure}

% Introduce the problem, the data and the model
We follow \cite{Charpentier2015} to construct a copula model to capture the dependency structure between prices of oil future contracts. The data considered is presented in Figure~\ref{fig:example3-oil-copula-data} and consists of weekly log-returns between January 10, 1997 and June 4, 2010 of Brent (produced in the North Sea), Dubai (produced in the Persian Gulf) and Maya (produced in the Gulf of Mexico) oil. We partition the data set into two parts and make use of the first $465$ data points for estimating the $\alpha$\textsc{sv} and copula models. The remaining $233$ data points are kept for validating the model by backtesting.
 
\begin{algorithm}[!t]
\footnotesize
\caption{\textsf{Copula modelling using $\alpha$\textsc{sv} as the marginal models}}
\textsf{Stage 1} (repeated for each asset $i$)
\begin{algorithmic}[1]
	\STATE  Run Algorithm~\ref{alg:GPOABC} to obtain the log-volatility estimate $\widehat{x}_{1:T,i}$ and the parameter estimate $\widehat{\theta}_{\text{MAP},i}$ of \eqref{eq:aSVmodel4parameters}. Compute the filtered log-returns $\widehat{e}_{t,i}$ by
	\begin{align}
			\widehat{e}_{t,i} = \exp \bigg( - \frac{1}{2} \widehat{x}_{t,i} \bigg) y_t, \qquad \text{ for } t=1,\ldots,T.
			\label{eq:FilteredResiduals}
	\end{align}
	\STATE Estimate the cumulative distribution function (\textsc{cdf}) denoted by $\widehat{G}_{i}$ from $\{\widehat{e}_{t,i}\}_{t=1}^T$. Compute the probability transformation of the residuals by
	\begin{align}
		\widehat{u}_{t,i} = \widehat{G}_{i}(\widehat{e}_{t,i}), \qquad \text{ for } t=1,\ldots,T.
		\label{eq:TransformedFilteredResiduals}
	\end{align}
\end{algorithmic}	
\textsf{Stage 2}
\begin{algorithmic}[1]
	\setalglineno{4}
	\STATE Infer the parameters of the copula to model the dependency between $\{ \{ \widehat{u}_{t,i} \}_{t=1}^T \}_{i=1}^{30}$.
\end{algorithmic}
\label{alg:copula}
\end{algorithm}

% How to estimate the residuals and use the ECDF to obtain the uniforms for the copula
We adopt the commonly used two-stage approach to copula modelling outlined in Algorithm~\ref{alg:copula}, where marginal models are first fitted separately to each of the log-return series, and then combined using a copula to model the dependency structure \citep{Joe2005}. We make use of the procedure in Section~\ref{sec:results:alpha} to estimate the parameters of an $\alpha$\textsc{sv} model for each type of oil. The filtered residuals \eqref{eq:FilteredResiduals} are assumed to be independent and identically distributed as $\mathcal{A}(\widehat{\alpha}_i,1)$ by \eqref{eq:aSVmodel4parameters} if $x_{1:T}$ is known. Stage~1 is carried out independently for each asset and is therefore straight-forward to implemented in a parallel manner. The use of the proposed algorithm decreases the computational time for this stage from hours or days for each asset to about half an hour compared with \textsc{pmh}.

We follow \citet[p. 231-231]{McNeilFreyEmbrechts2010} to model the dependency in the residuals. This amounts to applying a probability transform via the empirical \textsc{cdf} on the residuals into the $3$-dimensional hypercube, see right part of Figure~\ref{fig:example3-oil-copula-data}. The transformed residuals $\{\widehat{u}_{1:T,1},\ldots,\widehat{u}_{1:T,d}\}$ are then combined by a Student's $t$-copula function to find a model for the joint distribution. The degrees of freedom and the correlation matrix in the Student's $t$-copula are estimated using \textsc{map} and matching of moments via Kendall's $\tau$, respectively.

\begin{figure}[t]
	\centering
	\includegraphics[width=\textwidth]{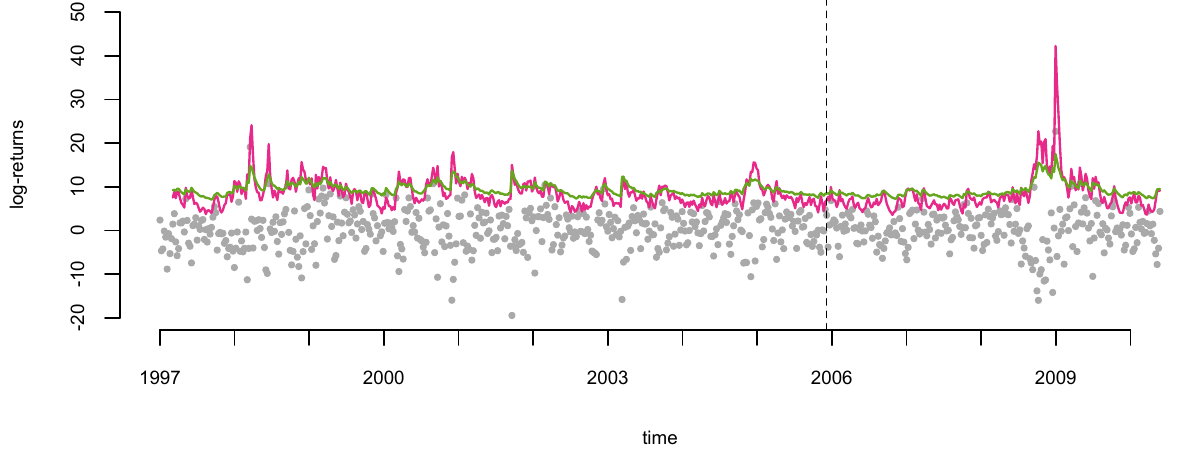}
	\caption{Estimated values of $\mathsf{VaR}_{0.99}(e_t)$ for an equally weighted portfolio of the three oil futures using the \textsc{gsv} (magenta) and $\alpha$\textsc{sv} (green) model with the Student-$t$ copula. The dashed line indicates the division of estimation and validation data.}
	\label{fig:example3-oil-copula-var}
\end{figure}

Finally, we make use of the copula models and their margins to estimate the \textsc{var} for each type of oil. The \textsc{var} at confidence $\bar{\alpha} \in (0,1)$ is defined by
\begin{align*}
	\mathsf{VaR}_{\bar{\alpha}}(e_t) = \inf \Big\{ -e_t \in \mathbb{R}: G(-e_t) \geq \bar{\alpha} \Big\},
\end{align*}
i.e.\ the smallest loss $-e_t$ such that probability that the loss (the negative log-return) exceeds $-e_t$ is no larger then $(1-\bar{\alpha})$. We adopt a Monte Carlo approach to estimate $\mathsf{VaR}_{\bar{\alpha}}(e_t)$ by: (i) simulating from the copula, (ii) obtain simulated filtered residuals by applying a quantile transform based on the empirical \textsc{cdf}, (iii) computing the resulting log-returns by the estimated volatility and the inverse of \eqref{eq:FilteredResiduals}. We repeat this $100,000$ times for each asset and then compute the empirical quantile corresponding to $\mathsf{VaR}_{0.99}(e_t)$ for an equally weighted portfolio. However, more advanced portfolio weighting schemes can be easily implemented.

The resulting \textsc{var}-estimate is presented Figure~\ref{fig:example3-oil-copula-var}, where we compared the $\alpha$\textsc{sv} model with a \textsc{gsv} model \eqref{eq:SVmodel3parameters} estimated using \textsc{gs} in a similar manner. We note that the estimates from the two models are quite different especially at the end of the data series. Backtesting on the validation data gives $0$ violations for both models and the expected number of violations is $2.3$. The \textsc{gsa} algorithm requires around $30-60$ minutes to infer the model for each of the three assets, where the inference is straight-forward to run on parallel architecture. The corresponding computational time required by the \textsc{pmh} algorithm would be in the order of $18-24$ hours.

\subsection{Computing VAR for a portfolio of stocks}
\label{sec:results:application:stocks}
We offer a final numerical example to illustrate that the proposed method can be applied to large portfolios as well. The data that we consider is the 30 industrial portfolio provided by Kenneth French. The portfolio consists of monthly log-returns of $30$ different industrial sectors during the period September, 1926 to December, 2015. Again, we partition the data into an estimation set and a validation set with $716$ and $358$ data points, respectively.

\begin{figure}[t]
	\centering
	\includegraphics[width=\textwidth]{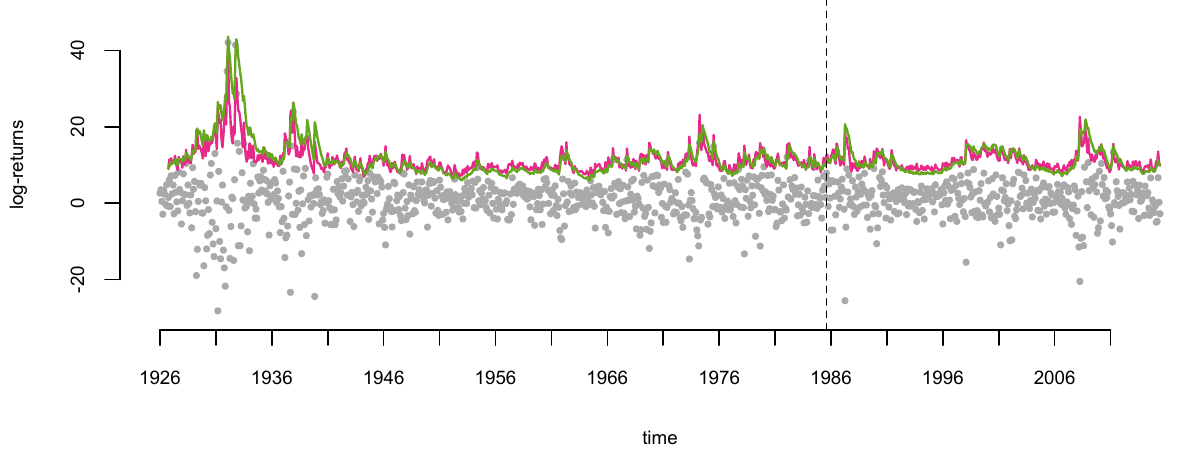}
	\caption{The log-returns (grey dots) and the estimated values of $\mathsf{VaR}_{0.99}(e_t)$ for an equally weighted portfolio of stocks using the \textsc{gsv} (magenta) and $\alpha$\textsc{sv} (green) model with the Student's $t$-copula. The dashed line indicates the division of estimation and validation data.}
	\label{fig:example4-portfolio}
\end{figure}

We adopt the same procedure as in Section~\ref{sec:results:application:oil} and the results are presented in Figure~\ref{fig:example4-portfolio}. The conclusions are similar with both \textsc{var} estimates being quite similar. Backtesting gives $0$ and $1$ violations for the $\alpha$\textsc{sv} and \textsc{gsv} models, respectively. The expected number of violations is $3.5$. The computational speed-up with \textsc{gsa} compared with \textsc{pmh} is similar to in Section~\ref{sec:results:application:oil} as the number of observations per asset are similar.

%%%%%%%%%%%%%%%%%%%%%%%%%%%%%%%%%%%%%%%%%%%%%%%%%%%%%%%%%%%%%%%%%%%%%%%%%%%%%%%%%%%%%%%%%%%%%%%%%%%%%%
%%%%%%%%%%%%%%%%%%%%%%%%%%%%%%%%%%%%%%%%%%%%%%%%%%%%%%%%%%%%%%%%%%%%%%%%%%%%%%%%%%%%%%%%%%%%%%%%%%%%%%%
%%%%%%%%%%%%%%%%%%%%%%%%%%%%%%%%%%%%%%%%%%%%%%%%%%%%%%%%%%%%%%%%%%%%%%%%%%%%%%%%%%%%%%%%%%%%%%%%%%%%%%%
\section{Conclusions}
\label{sec:conclusions}
We have proposed \textsc{gsa}, an algorithm to approximate the posterior distribution in \textsc{ssm}s with intractable likelihoods. The illustrations provided in Section~\ref{sec:results} indicate that the proposed algorithm is quite accurate and exhibits a substantial decrease in the computational cost. We obtain similar posterior estimates to \textsc{pmh} with a speed-up of one or two orders of magnitude, which reduces computational time from tens of hours to tens of minutes. Moreover, \textsc{gsa} seems to be quite robust to noise in the log-posterior estimates, which typically results in that the \textsc{pmh} algorithm gets stuck at times and that the \textsc{spsa} algorithm converges slowly. Overall, this shows that the proposed algorithm is an efficient inference method that makes it possible for practitioners to use models with intractable likelihoods, such as copula models with $\alpha$-stable margins, in applied work.

Future work includes: (i) adopting a sparse representation of the \textsc{gp}, (ii) developing new acqusition functions, (iii) making use of better tailored covariance functions and (iv) incorperating adaptive approaches for choosing the tolerance level $\epsilon$. Some ideas for sequential and sparse representations of \textsc{gp}s are discussed by \cite{Huber2014} and \cite{BijlWingerdenSchonVerhaegen2015}. It would also be interesting to design new acquisition functions to obtain good estimates of the Hessian of the log-posterior or higher order moments at the same time as the \textsc{map} estimate. This could improve the Laplace approximation of the parameter posterior and open up for alternative posterior approximations such as using the skewed Student's $t$-distribution. Moreover, an interesting approach for tailored \textsc{gp} priors would be to make use of non-stationary covariance functions \citep{PaciorekSchervish2004}. This would capture the fact that the log-posterior often falls off rapidly in some parts of the parameter space and is almost flat in other parts. 

Finally, it would be beneficial to include some adaptive approach to select $\epsilon$ in the \textsc{smc}-\textsc{abc} algorithm to decrease the number of choices for the user. As previously discussed, we initially implemented the adaptive algorithms proposed in \cite{DelMoralDoucetJasra2012} and \cite{CalvetCzellar2015}. However, we ran into problems when using them to approximate the log-posterior values using moderate values of $N$. The estimates exhibited a large bias that resulted in biased estimates of $\theta$ using \textsc{pmh}. As a result, we choose not to adapt $\epsilon$ to be able to keep $N$ much smaller. It is therefore interesting to explore other adaptive schemes that provide good estimates of the log-posterior using a moderate value of $N$.

%%%%%%%%%%%%%%%%%%%%%%%%%%%%%%%%%%%%%%%%%%%%%%%%%%%%%%%%%%%%%%%%%%%%%%%%%%%%%%%%%%%%%%%%%%%%%%%%%%%%%%%
%%%%%%%%%%%%%%%%%%%%%%%%%%%%%%%%%%%%%%%%%%%%%%%%%%%%%%%%%%%%%%%%%%%%%%%%%%%%%%%%%%%%%%%%%%%%%%%%%%%%%%%
%%%%%%%%%%%%%%%%%%%%%%%%%%%%%%%%%%%%%%%%%%%%%%%%%%%%%%%%%%%%%%%%%%%%%%%%%%%%%%%%%%%%%%%%%%%%%%%%%%%%%%%
\section*{Acknowledgements}
This work was supported by: \textit{Probabilistic modeling of dynamical systems} (Contract number: 621-2013-5524) and \textsc{cadics}, a Linnaeus Center, both funded by the Swedish Research Council. The simulations were performed on resources provided by the Swedish National Infrastructure for Computing (\textsc{snic}) at Link\"{o}ping University, Sweden. J.\ Dahlin would like to thank Fredrik Lindsten, Neil Lawrence and Carl-Henrik Ek for fruitful discussions. Thanks also to Joerg M.\ Gablonsky, Abraham Lee, Per A. Brodtkorb and the GPy team for making their Python implementations available.

%%%%%%%%%%%%%%%%%%%%%%%%%%%%%%%%%%%%%%%%%%%%%%%%%%%%%%%%%%%%%%%%%%%%%%%%%%%%%%%%%%%%%%%%%%%%%%%%%%%%%%%
%%%%%%%%%%%%%%%%%%%%%%%%%%%%%%%%%%%%%%%%%%%%%%%%%%%%%%%%%%%%%%%%%%%%%%%%%%%%%%%%%%%%%%%%%%%%%%%%%%%%%%%
%%%%%%%%%%%%%%%%%%%%%%%%%%%%%%%%%%%%%%%%%%%%%%%%%%%%%%%%%%%%%%%%%%%%%%%%%%%%%%%%%%%%%%%%%%%%%%%%%%%%%%%
\appendix
\section{Implementation details}
\label{app:implementationdetails}
\textit{\textsc{gpo} algorithm:} We make use of the GPy package \citep{gpy2014} for calculating the \textsc{gp} predictive posterior and estimating the \textsc{gp} prior hyperparameters. In this paper, we assume a zero prior mean function $m(\theta)=0$ and a combination of the bias and the Mat\'{e}rn 5/2 covariance functions. This choice corresponds to a prior for the log-posterior with some non-zero mean and two continuous derivatives. These are reasonable assumptions as this kind of smoothness is assumed in the Laplace approximation. We estimate the hyperparameters using \textit{empirical Bayes} (\textsc{eb}), i.e.\ by optimising the marginal likelihood with respect to $\lambda$. More advanced schemes that marginalise over the hyperparameters using slice sampling \citep{MurrayAdamsMacKay2010} or \textsc{smc} \citep{SvenssonDahlinSchon2015} can be used within \textsc{sga} as well. 

For the acquisition rule in \eqref{eq:gpo:aqoptimisation}, we use $\zeta=0.01$ and $\Sigma=0.01 \mathbf{I}_p$. We initialise the \textsc{gpo} algorithm using $L=50$ samples obtained using Latin hypercube sampling with the implementation written by Abraham Lee available at \url{https://pypi.python.org/pypi/pyDOE}. The optimisation problems in Lines~8 and 11 in Algorithm~\ref{alg:GPOABC} are solved using the \textsc{direct} implementation written by Joerg M.\ Gablonsky, available from \url{https://pypi.python.org/pypi/DIRECT/}. Finally for Line~12, we make use of the Python implementation by Per A. Brodtkorb available at \url{https://pypi.python.org/pypi/Numdifftools}.

\textit{Section~\ref{sec:results:Gaussian}:} We use $N=2,000$ particles in \textsc{gs} and $N=2,000$ particles with the Gaussian density with tolerance level $\epsilon=0.20$ and $\psi(x)=x$ in \textsc{gsa} to produce the results in Figure~\ref{fig:example1-posteriors}.  We run the \textsc{gpo} algorithms for $K=450$ iterations after the initialisation and re-estimate the hyperparameters of the \textsc{gp} prior every $25$th iteration. The search space for the \textsc{gpo} algorithm $\Theta_{\text{GPO}}$ is given by $\mu \in (0,1)$, $\phi \in (0,1)$ and $\sigma_v \in (0.01,1)$. We use the following prior densities
\begin{align*}
	p(\mu)      \sim \mathcal{N}(\mu;0,0.2^2), \quad
	p(\phi)     \sim \mathcal{TN}_{(-1,1)}(\phi;0.9,0.05^2), \quad
	p(\sigma_v) \sim \mathcal{G}(\sigma_v;2,20),
\end{align*}
where $\mathcal{TN}_{(a,b)}(\cdot)$ denotes a truncated Gaussian distribution on $[a,b]$, $\mathcal{G}(a,b)$ denotes the Gamma distribution with mean $a/b$.

For \textsc{pmh}, we make use of the \textsc{smc}-\textsc{abc} algorithm with $N=2,000$ particles to estimate the log-posterior. We initialise \textsc{pmh} in $\theta_0=\{0.10,0.95,0.12\}$ and run the algorithm for $M=15,000$ iterations (discarding the first $5,000$ as burn-in). The parameter proposal is selected as 
\begin{align*}
	q(\theta'|\theta_k) 
	= 
	\mathcal{N}(\theta'; \theta_k, \Sigma_q),
	\qquad
	\Sigma_q
	=
	\frac{2.562^2}{3}
	\cdot
	10^{-4}
	\cdot
	\mathsf{diag}(137, 7, 38),
\end{align*}
which results from an asymptotic rule-of-thumb, see \cite{DahlinSchon2015}, with an estimate of the posterior covariance from a pilot run.

For \textsc{spsa}, we make use of $N=2,000$ particles and follow \cite{Spall1998} to select the hyperparameters as $a=0.001$, $c=0.30$, $A=35$, $\alpha=0.602$ and $\gamma=0.101$ using pilot runs.

\textit{Section~\ref{sec:results:alpha}:} The real-world data is computed as $y_t = 100 [ \log(s_{t}) - \log(s_{t-1}) ]$ , where $s_t$ denotes the price of a future contract on coffee\footnote{Data available at: \url{https://www.quandl.com/CHRIS/ICE_KC2}.}. We follow \cite{YildirimSinghDeanJasra2014} and apply $\psi(x)=\arctan(x)$ to stabilise the variance of the likelihood (and gradient estimate). A two step approach is applied to sample from the zero-mean symmetric $\alpha$-stable distribution $\mathcal{A}(\alpha,\gamma)$. First, we sample $v_t^{(1)} \sim \textsf{Exp}(1)$ and $v_t^{(2)} \sim \mathcal{U}(-\pi/2,\pi/2)$. Then, we obtain a sample (when $\alpha \neq 1$) by applying the transformation
\begin{align*}
\check{y}_t &= 
	\gamma
	\frac{ \sin \big( \alpha v^{(2)}_t \big) }
	{\big[ \cos (v^{(2)}_t) \big]^{1/\alpha}}
	\Bigg[
	\frac
	{\cos \big[ (\alpha - 1) v^{(2)}_t \big]}{v^{(1)}_t}
	\Bigg]^{\frac{1-\alpha}{\alpha}}.
\end{align*}
See \cite{Nolan2003} for more on the generation of $\alpha$-stable random numbers. 

We use $N=2,000$ particles with the Gaussian density with tolerance level $\epsilon=0.10$ in \textsc{smc}-\textsc{abc} to estimate the log-posterior. We run the \textsc{gpo} algorithm using the same settings as before but add $\alpha \in (1.2,2)$ to the search space and $p(\alpha) \sim \mathcal{B}(\alpha/2;20,2)$ to the prior distributions, where $\mathcal{B}(a,b)$ denotes the Beta distribution. We initialise \textsc{pmh} in $\theta_0=\{0.22,0.93,0.25,1.55\}$ and the parameter proposal is selected using a pilot run as 
\begin{align*}
	q(\theta'|\theta_k) 
	= 
	\mathcal{N}(\theta'; \theta_k, \Sigma_q),
	\qquad
	\Sigma_q
	=
	\frac{2.562^2}{4}
	\cdot
	10^{-3}
	\cdot
	\mathsf{diag}(26, 1, 9, 11).
\end{align*}

\textit{Section~\ref{sec:results:application:oil} and \ref{sec:results:application:stocks}:} Most of the settings are the same for the oil\footnote{Data available at: \url{http://freakonometrics.free.fr/oil.xls}.} and stock\footnote{Data available at: \url{http://mba.tuck.dartmouth.edu/pages/faculty/ken.french/data_library.html}.} portfolio examples. We use $N=5,000$ particles in the \textsc{smc} and \textsc{smc}-\textsc{abc} algorithms and keep the remaining settings as Sections~\ref{sec:results:Gaussian} and \ref{sec:results:alpha}. For the stock portfolio example, we change the search space of $\mu$ to $(0,4)$ as the weekly log-returns in this data set can be much larger than the daily log-returns in the other data sets. The \textsc{map} estimate of the degrees of freedom in the copula is obtained by a quasi-Newton solver using a uniform prior.

%%%%%%%%%%%%%%%%%%%%%%%%%%%%%%%%%%%%%%%%%%%%%%%%%%%%%%%%%%%%%%%%%%%%%%%%%%%%%%%%%%%%%%%%%%%%%%%%%%%%%%%
%%%%%%%%%%%%%%%%%%%%%%%%%%%%%%%%%%%%%%%%%%%%%%%%%%%%%%%%%%%%%%%%%%%%%%%%%%%%%%%%%%%%%%%%%%%%%%%%%%%%%%%
%%%%%%%%%%%%%%%%%%%%%%%%%%%%%%%%%%%%%%%%%%%%%%%%%%%%%%%%%%%%%%%%%%%%%%%%%%%%%%%%%%%%%%%%%%%%%%%%%%%%%%%
\bibliographystyle{plainnat}
\bibliography{dahlin}

\begin{thebibliography}{36}
\providecommand{\natexlab}[1]{#1}
\providecommand{\url}[1]{\texttt{#1}}
\expandafter\ifx\csname urlstyle\endcsname\relax
  \providecommand{\doi}[1]{doi: #1}\else
  \providecommand{\doi}{doi: \begingroup \urlstyle{rm}\Url}\fi

\bibitem[Andrieu et~al.(2010)Andrieu, Doucet, and
  Holenstein]{AndrieuDoucetHolenstein2010}
C.~Andrieu, A.~Doucet, and R.~Holenstein.
\newblock {Particle Markov chain Monte Carlo methods}.
\newblock \emph{Journal of the Royal Statistical Society: Series B (Statistical
  Methodology)}, 72\penalty0 (3):\penalty0 269--342, 2010.

\bibitem[Bijl et~al.(2015)Bijl, van Wingerden, Sch\"on, and
  Verhaegen]{BijlWingerdenSchonVerhaegen2015}
H.~Bijl, J-W. van Wingerden, T.~B. Sch\"on, and M.~Verhaegen.
\newblock {Online sparse Gaussian process regression using FITC and PITC
  approximations}.
\newblock In \emph{Proceedings of the 17th IFAC Symposium on System
  Identification (SYSID)}, pages 703--708, Beijing, China, October 2015.

\bibitem[Brochu et~al.(2010)Brochu, Cora, and
  De~Freitas]{BrochuCoraDeFreitas2010}
E.~Brochu, V.~M. Cora, and N.~De~Freitas.
\newblock {A tutorial on Bayesian optimization of expensive cost functions,
  with application to active user modeling and hierarchical reinforcement
  learning}.
\newblock \emph{Pre-print}, 2010.
\newblock arXiv:1012.2599v1.

\bibitem[Bull(2011)]{Bull2011}
A.~D. Bull.
\newblock Convergence rates of efficient global optimization algorithms.
\newblock \emph{Journal of Machine Learning Research}, 12:\penalty0 2879--2904,
  2011.

\bibitem[Calvet and Czellar(2015)]{CalvetCzellar2015}
L.~E. Calvet and V.~Czellar.
\newblock Accurate methods for approximate {B}ayesian computation filtering.
\newblock \emph{Journal of Financial Econometrics}, 13\penalty0 (4):\penalty0
  798--838, 2015.

\bibitem[Charpentier(2015)]{Charpentier2015}
A.~Charpentier.
\newblock {Pr{\'e}vision avec des copules en finance}.
\newblock Technical report, May 2015.
\newblock URL \url{https://hal.archives-ouvertes.fr/hal-01151233}.

\bibitem[Dahlin and Lindsten(2014)]{DahlinLindsten2014}
J.~Dahlin and F.~Lindsten.
\newblock {Particle filter-based Gaussian process optimisation for parameter
  inference}.
\newblock In \emph{Proceedings of the 19th IFAC World Congress}, Cape Town,
  South Africa, August 2014.

\bibitem[Dahlin and Sch\"{o}n(2015)]{DahlinSchon2015}
J.~Dahlin and T.~B. Sch\"{o}n.
\newblock {Getting started with particle Metropolis-Hastings for inference in
  nonlinear models}.
\newblock \emph{Pre-print:}, 2015.
\newblock { arXiv:1511.01707v4}.

\bibitem[Dean and Singh(2011)]{DeanSingh2011}
T.~A. Dean and S.~S. Singh.
\newblock Asymptotic behaviour of approximate {B}ayesian estimators.
\newblock \emph{Pre-print:}, 2011.
\newblock { arXiv:1105.3655v1}.

\bibitem[Del~Moral et~al.(2012)Del~Moral, Doucet, and
  Jasra]{DelMoralDoucetJasra2012}
P.~Del~Moral, A.~Doucet, and A.~Jasra.
\newblock {An adaptive sequential Monte Carlo method for approximate Bayesian
  computation}.
\newblock \emph{Statistics and Computing}, 22\penalty0 (5):\penalty0
  1009--1020, 2012.

\bibitem[Doucet and Johansen(2011)]{DoucetJohansen2011}
A.~Doucet and A.~Johansen.
\newblock A tutorial on particle filtering and smoothing: Fifteen years later.
\newblock In D.~Crisan and B.~Rozovsky, editors, \emph{The Oxford Handbook of
  Nonlinear Filtering}. Oxford University Press, 2011.

\bibitem[Durbin and Koopman(2012)]{DurbinKoopman2012}
J.~Durbin and S.~J. Koopman.
\newblock \emph{Time series analysis by state space methods}.
\newblock Oxford University Press, 2 edition, 2012.

\bibitem[Ehrlich et~al.(2015)Ehrlich, Jasra, and
  Kantas]{EhrlichJasraKantas2015}
E.~Ehrlich, A.~Jasra, and N.~Kantas.
\newblock Gradient free parameter estimation for hidden {M}arkov models with
  intractable likelihoods.
\newblock \emph{Methodology and Computing in Applied Probability}, 17\penalty0
  (2):\penalty0 315--349, 2015.

\bibitem[Gutmann and Corander(2016)]{GutmannCorander2015}
M.~U. Gutmann and J.~Corander.
\newblock Bayesian optimization for likelihood-free inference of
  simulator-based statistical models.
\newblock \emph{Journal of Machine Learning Research}, 17\penalty0
  (125):\penalty0 1--47, 2016.

\bibitem[Huber(2014)]{Huber2014}
M.~F. Huber.
\newblock {Recursive Gaussian process: On-line regression and learning}.
\newblock \emph{Pattern Recognition Letters}, 45:\penalty0 85--91, 2014.

\bibitem[Jasra et~al.(2012)Jasra, Singh, Martin, and
  McCoy]{JasraSinghMartinMcCoy2012}
A.~Jasra, S.~S. Singh, J.~S. Martin, and E.~McCoy.
\newblock {Filtering via approximate Bayesian computation}.
\newblock \emph{Statistics and Computing}, 22\penalty0 (6):\penalty0
  1223--1237, 2012.

\bibitem[Joe(2005)]{Joe2005}
H.~Joe.
\newblock Asymptotic efficiency of the two-stage estimation method for
  copula-based models.
\newblock \emph{Journal of Multivariate Analysis}, 94\penalty0 (2):\penalty0
  401--419, 2005.

\bibitem[Jones et~al.(1993)Jones, Perttunen, and Stuckman]{Jones1993}
D.~R. Jones, C.~D. Perttunen, and B.~E. Stuckman.
\newblock {Lipschitzian optimization without the Lipschitz constant}.
\newblock \emph{{Journal of Optimization Theory and Applications}}, 79\penalty0
  (1):\penalty0 157--181, 1993.

\bibitem[Lizotte(2008)]{Lizotte2008}
D.~J. Lizotte.
\newblock \emph{{Practical Bayesian optimization}}.
\newblock PhD thesis, University of Alberta, 2008.

\bibitem[Ljung(1999)]{Ljung1999}
L.~Ljung.
\newblock \emph{System identification: theory for the user}.
\newblock Prentice Hall, 1999.

\bibitem[Marin et~al.(2012)Marin, Pudlo, Robert, and
  Ryder]{MarinPudloRobertRyder2012}
J-M. Marin, P.~Pudlo, C.~P. Robert, and R.~J. Ryder.
\newblock {Approximate Bayesian computational methods}.
\newblock \emph{Statistics and Computing}, 22\penalty0 (6):\penalty0
  1167--1180, 2012.

\bibitem[McNeil et~al.(2010)McNeil, Frey, and
  Embrechts]{McNeilFreyEmbrechts2010}
A.~J. McNeil, R.~Frey, and P.~Embrechts.
\newblock \emph{Quantitative risk management: concepts, techniques, and tools}.
\newblock Princeton University Press, 2010.

\bibitem[Meeds and Welling(2014)]{MeedsWelling2014}
E.~Meeds and M.~Welling.
\newblock {GPS-ABC}: {G}aussian process surrogate approximate {B}ayesian
  computation.
\newblock In \emph{Proceedings of the 30th Conference on Uncertainty in
  Artificial Intelligence (UAI)}, Quebec City, Canada, July 2014.

\bibitem[Murray et~al.(2010)Murray, Adams, and MacKay]{MurrayAdamsMacKay2010}
I.~Murray, R.~Adams, and D.~MacKay.
\newblock Elliptical slice sampling.
\newblock In \emph{Proceedings of the 13th International Conference on
  Artificial Intelligence and Statistics (AISTATS)}, pages 541--548, Sardinia,
  Italy, May 2010.

\bibitem[Nolan(2003)]{Nolan2003}
J.~Nolan.
\newblock \emph{Stable distributions: models for heavy-tailed data}.
\newblock Birkhauser, 2003.

\bibitem[Paciorek and Schervish(2004)]{PaciorekSchervish2004}
C.~J. Paciorek and M.~J. Schervish.
\newblock Nonstationary covariance functions for {G}aussian process regression.
\newblock In \emph{Proceedings of the 2004 Conference on Neural Information
  Processing Systems ({NIPS})}, pages 273--280, Vancouver, Canada, December
  2004.

\bibitem[Panov and Spokoiny(2015)]{PanovSpokoiny2015}
M.~Panov and V.~Spokoiny.
\newblock Finite sample {B}ernstein-von-{M}ises theorem for semiparametric
  problems.
\newblock \emph{Bayesian Analysis}, 10\penalty0 (3):\penalty0 665--710, 2015.

\bibitem[Pitt et~al.(2012)Pitt, Silva, Giordani, and
  Kohn]{PittSilvaGiordaniKohn2012}
M.~K. Pitt, R.~S. Silva, P.~Giordani, and R.~Kohn.
\newblock On some properties of {M}arkov chain {M}onte {C}arlo simulation
  methods based on the particle filter.
\newblock \emph{Journal of Econometrics}, 171\penalty0 (2):\penalty0 134--151,
  2012.

\bibitem[Rasmussen and Williams(2006)]{RasmussenWilliams2006}
C.~E. Rasmussen and C.~K.~I. Williams.
\newblock \emph{Gaussian processes for machine learning}.
\newblock {MIT} Press, 2006.

\bibitem[Spall(1998)]{Spall1998}
J.~C. Spall.
\newblock Implementation of the simultaneous perturbation algorithm for
  stochastic optimization.
\newblock \emph{IEEE Transactions on Aerospace and Electronic Systems},
  34\penalty0 (3):\penalty0 817--823, 1998.

\bibitem[Stoyanov et~al.(2010)Stoyanov, Racheva-Iotova, Rachev, and
  Fabozzi]{StoyanovRachevaRachevFabozzi2010}
S.~V. Stoyanov, B.~Racheva-Iotova, S.~T. Rachev, and F.~J. Fabozzi.
\newblock Stochastic models for risk estimation in volatile markets: a survey.
\newblock \emph{Annals of Operations Research}, 176\penalty0 (1):\penalty0
  293--309, 2010.

\bibitem[Svensson et~al.(2015)Svensson, Dahlin, and
  Sch\"on]{SvenssonDahlinSchon2015}
A.~Svensson, J.~Dahlin, and T.~B. Sch\"on.
\newblock {Marginalizing Gaussian process hyperparameters using sequential
  Monte Carlo}.
\newblock In \emph{Proceedings of the 6th IEEE International Workshop on
  Computational Advances in Multi-Sensor Adaptive Processing (CAMSAP)}, Cancun,
  Mexico, December 2015.

\bibitem[{The GPy authors}(2014)]{gpy2014}
{The GPy authors}.
\newblock {GPy}: A {G}aussian process framework in {P}ython.
\newblock \url{http://github.com/SheffieldML/GPy}, 2014.

\bibitem[Vazquez and Bect(2010)]{VazquezBect2010}
E.~Vazquez and J.~Bect.
\newblock Convergence properties of the expected improvement algorithm with
  fixed mean and covariance functions.
\newblock \emph{Journal of Statistical Planning and inference}, 140\penalty0
  (11):\penalty0 3088--3095, 2010.

\bibitem[Wood(2010)]{Wood2010}
S.~N. Wood.
\newblock Statistical inference for noisy nonlinear ecological dynamic systems.
\newblock \emph{Nature Letters}, 466:\penalty0 1102--1104, 2010.

\bibitem[Y{\i}ld{\i}r{\i}m et~al.(2014)Y{\i}ld{\i}r{\i}m, Singh, Dean, and
  Jasra]{YildirimSinghDeanJasra2014}
S.~Y{\i}ld{\i}r{\i}m, S.~S. Singh, T.~Dean, and A.~Jasra.
\newblock {Parameter estimation in hidden Markov models with intractable
  likelihoods using sequential Monte Carlo}.
\newblock \emph{Journal of Computational and Graphical Statistics}, 24\penalty0
  (3):\penalty0 846--865, 2014.

\end{thebibliography}

\end{document}